\documentclass[12pt,manuscript]{aastex}
\usepackage{apjfonts}

\newcommand{\msun}{M_\odot}

\newcommand{\gapprox}{\mathrel{\mathpalette\@versim>}}
\newcommand{\lapprox}{\mathrel{\mathpalette\@versim<}}
\newcommand{\propapprox}{\mathrel{\mathpalette\@versim\propto}}

\shorttitle{Ejecta expansion in Tycho} 
\shortauthors{WILLIAMS ET AL.}

\begin{document}

\title{The Three-Dimensional Expansion of the Ejecta from Tycho's
  Supernova Remnant}

\author{Brian J. Williams,\altaffilmark{1,2}
Nina M. Coyle,\altaffilmark{2,3}
Hiroya Yamaguchi,\altaffilmark{2,4}
Joseph Depasquale,\altaffilmark{1,5}
Ivo R. Seitenzahl,\altaffilmark{6,7}
John W. Hewitt,\altaffilmark{8}
John M. Blondin,\altaffilmark{9}
Kazimierz J. Borkowski,\altaffilmark{9}
Parviz Ghavamian,\altaffilmark{10}
Robert Petre,\altaffilmark{2}
Stephen P. Reynolds,\altaffilmark{9}
}

\altaffiltext{1}{Space Telescope Science Institute, Baltimore, MD 21218; bwilliams@stsci.edu}
\altaffiltext{2}{NASA Goddard Space Flight Center, X-ray Astrophysics Laboratory, Greenbelt, MD 20771, USA}
\altaffiltext{3}{University of Chicago, Department of Physics, Chicago, Illinois 60637}
\altaffiltext{4}{CRESST/University of Maryland, College Park, College Park, MD 20742}
\altaffiltext{5}{Harvard-Smithsonian Center for Astrophysics, Cambridge, MA 02138}
\altaffiltext{6}{School of Physical, Environmental and Mathematical Sciences, University of New South Wales, Australian Defence Force Academy, Canberra, ACT 2600, Australia}
\altaffiltext{7}{Research School of Astronomy and Astrophysics, Australian National University, Canberra, ACT 2611, Australia}
\altaffiltext{8}{University of North Florida, Department of Physics, 1 UNF Drive, Jacksonville, FL 32224, USA}
\altaffiltext{9}{Department of Physics, North Carolina State University, Raleigh, NC 27695}
\altaffiltext{10}{Department of Physics, Astronomy, and Geosciences, Towson University, Towson, MD 21252}

\begin{abstract}

We present the first three-dimensional measurements of the velocity of
various ejecta knots in Tycho's supernova remnant, known to result
from a Type Ia explosion. Chandra X-ray observations over a 12-year
baseline from 2003 to 2015 allow us to measure the proper motion of
nearly 60 ``tufts'' of Si-rich ejecta, giving us the velocity in the
plane of the sky. For the line of sight velocity, we use two different
methods: a non-equilibrium ionization model fit to the strong Si and S
lines in the 1.2-2.8 keV regime, and a fit consisting of a series of
Gaussian lines. These methods give consistent results, allowing us to
determine the red or blue shift of each of the knots. Assuming a
distance of 3.5 kpc, we find total velocities that range from 2400 to
6600 km s$^{-1}$, with a mean of 4430 km s$^{-1}$. We find several
regions where the ejecta knots have overtaken the forward shock. These
regions have proper motions in excess of 6000 km s$^{-1}$. Some Type
Ia supernova explosion models predict a velocity asymmetry in the
ejecta. We find no such velocity asymmetries in Tycho, and discuss our
findings in light of various explosion models, favoring those delayed
detonation models with relatively vigorous and symmetrical
deflagrations. Finally, we compare measurements with models of the
remnant's evolution that include both smooth and clumpy ejecta
profiles, finding that both ejecta profiles can be accommodated by
the observations.


\end{abstract}

\section{Introduction}
\label{intro}

Tycho's supernova remnant (SNR; hereafter Tycho) is the remnant of the
supernova (SN) first observed in 1572 CE \citep{stephenson02}. It was
classified by \citet{baade45} as a ``Type I'' event, and an analysis
of the X-ray emitting ejecta suggested a ``normal'' Typa Ia SN
event \citep{badenes06}. This was confirmed via detection and spectroscopy of light
echoes by \citet{rest08} and \citet{krause08}. These SNe are generally
believed to result from a thermonuclear explosion of a white dwarf in
a binary system, destabilized by mass transfer. While the nature of
the binary companion is unclear, there are two leading scenarios: the
single-degenerate (SD) model, in which a white dwarf accretes matter
from a non-degenerate companion, exploding when it reaches the
Chandrasekhar limit of $\sim 1.4$ $\msun$ \citep{whelan73}; and the
broadly defined double-degenerate (DD) model, which consists of an
explosion triggered by the merging of two white dwarfs by various
means \citep{webbink84}.

In an earlier work (\citealt{williams13}, hereafter ``Paper I''), we
examined {\it Spitzer} infrared (IR) observations of the remnant,
which show emission from interstellar dust grains, warmed in the
post-shock environment by collisions with energetic electrons and
ions. We fit models to the IR colors that allowed us to determine the
post-shock gas density, which we found to vary as a function of
azimuthal angle around the shell, with densities in the east and
northeast higher by a factor of several than those in the west and
southwest.

In a subsequent work (\citealt{williams16}, hereafter ``Paper II''),
we examined the proper motions of the forward shock in both X-rays and
radio. The emission processes in these two wavebands are the same:
nonthermal synchrotron radiation resulting from relativistic
electrons, accelerated by the amplified magnetic fields
\citep{ressler14,tran15} in the forward shock wave from the
supernova. We used {\it Karl G. Jansky Very Large Array (VLA)} radio
observations spread over 30 years and {\it Chandra} X-ray observations
spread over 15 years to measure the expansion rate of the remnant at
$\sim 20$ locations around the shell, finding that the velocity of the
forward shock varies by roughly a factor of two from one side of the
shell to the other. The direction of the velocity asymmetry is such
that the fastest shocks propagate into the lowest density environments
determined in Paper I, as expected.

The simplest explanation for this is that Tycho's SNR is expanding
into a non-uniform interstellar medium (ISM), such as a pre-existing
density gradient. However, another intriguing possibility is that the
explosion itself was non-uniform. The explosion mechanism for Type Ia
SNe is poorly understood, but there have been hints that at least some
explosions may be asymmetric. Supernovae themselves are unresolvable,
but spectroscopic information embedded in the emission lines during
the nebular phase has shown evidence in some SNe Ia for the red and
blueshifted ejecta velocities to be different
\citep{motohara06,maeda10}. \citet{wang08} summarize evidence for 
optical polarization in spectral lines in some SNe Ia before maximum, indicating
significant asymmetry. From the SNR side, there is evidence for
an asymmetric distribution of ejecta in the Type Ia SNRs G1.9+0.3
\citep{borkowski17} and SN 1006 \citep{winkler14}, but these are based
only on the spatial distribution, not the dynamic motions of the
ejecta. Although there is evidence for local asymmetricity in Tycho
\citep{yamaguchi17}, the global ejecta distribution has not been
investigated.

{\bf Previous studies by \citet{furuzawa09} and 
\citet{hayato10} examined the doppler broadening of the Fe and intermediate-mass
elements within the ejecta, finding that the Si, S, and Ar expand
at a higher velocity than the Fe. \citet{katsuda10} examined the proper motions
of the outer ejecta in five large ($\sim$ arcminute scale) regions around the 
periphery of Tycho, finding an average expansion rate of 0.294$''$ yr$^{-1}$.} In a recent paper by \citealt{sato16}, hereafter SH16, the authors
looked at several ``blobs'' of ejecta in Tycho's SNR, identifying both
red and blueshifted velocity components from the Doppler shifts of the
spectral lines in a deep (750 ks) 2009 {\it Chandra} observation of
Tycho. They find Doppler velocities in excess of 5000 km s$^{-1}$
along the line of sight for ejecta blobs near the center of the
remnant moving both towards and away from the observer. Their work
represents the first direct ejecta velocity measurements along a line of sight
direction in Tycho's SNR. 

In this work, we build upon the work of SH16 by extending the ejecta
velocity measurements to all three dimensions, building up a
velocity-vector map of nearly 60 spatially-coherent ejecta knots,
roughly evenly distributed throughout the remnant. The fluffy interior
of Tycho is dominated by emission from the ejecta, most prominently
the Si and S lines at $\sim 1.86$ and $\sim 2.45$ keV, respectively
\citep{warren05}. We combine line-of-sight (LOS) velocities measured
from the Doppler shifts of these spectral lines in each of these knots
with their proper motion in the plane of the sky, as measured from
observations in 2003 and 2015. With velocities in all three
dimensions, we derive absolute magnitudes and directions of the
velocity vectors of the ejecta knots, and show that the ejecta
velocities are consistent with a symmetric explosion. Similar work has
been done for core-collapse SNRs, such as Cas A
\citep{fesen06,delaney10}, but this work represents the first such map
of a Type Ia SNR.

\section{Observations}
\label{observations}

Tycho has been observed a total of five times with {\it Chandra}. A 50
ks observation in 2000 used the ACIS-S3 chip, which is not quite large
enough to fit the entire remnant on it. As a result, about 25\% of the
remnant, along the southern shell, is cut off by the chip edge
\citep{hwang02}. The next four observations (150 ks in 2003 and 2007,
750 ks in 2009, and 150 ks in 2015) all used the ACIS I-array and
cover the entire remnant. The choice of data sets that we use for this
work depends on whether we are measuring the proper motion or the LOS
velocity. For our proper motion measurements, we desire the longest
baseline possible with the same instrument, so we make our
measurements on the 2003 (PI: J.P. Hughes) and 2015 (PI:
B.J. Williams) data. Additional factors for this decision are that the
2003 exposure is much deeper than the 2000 exposure, the 2015
observation was specifically planned to match that from 2003, and the
2003 image covers the entire remnant, while the 2000 image cuts off
the southern portion of the remnant.

The 2003 observation was taken in a single pointing beginning on 2003
Apr 29, while the 2015 observation was also taken in a single pointing
beginning on 2015 Apr 22, a time baseline of 12.0 yr. We follow an
identical data reduction process as that described in Paper II, which
is based on the work of \citet{katsuda10}, where we use version 4.7 of
CIAO and version 4.6.5 of CALDB to process all epochs. We found no
significant background flaring in the light curves. We align all
epochs to a common reference frame (the deep 2009 observation is used
as the relative reference frame) using detected point sources in the
field of view. The images are slightly smoothed using a 2-pixel
Gaussian, which has virtually no effect on the profile shapes
described in Section~\ref{measurements}, but does significantly
decrease the pixel-to-pixel Poisson noise level.

Our spectral analysis of each region is performed entirely using the
2009 observation (PI: J.P. Hughes), which was split into nine
different segments between 2009 Apr 13 and 2009 May 3 with an
effective total exposure time of 734 ks. We follow a procedure used in
\citet{winkler14}, similar to that of SH16, stacking spectra for a
given region from all nine observations, using the {\it specextract}
tool in CIAO, weighting spectral files and response files
appropriately. Spectral fits were performed using XSPEC version
12.9.0p, which contains version 3.0 of AtomDB.

Our choice of knots for both spectral extraction and proper motion
measurement was guided by several requirements. First and most
importantly, we searched the entire remnant in both the 2003 and 2015
epochs for any structure that maintained a coherent shape between both
epochs {\em and} had a proper motion detectable by eye. This limited
us to regions away from the center of the remnant, as the ejecta
velocities there are mostly along the LOS. This is not a loss, though,
as these regions in the center of the remnant have been studied by
SH16. Secondly, we ensured that the knots were bright enough to get a
good signal-to-noise for the spectral fitting. This was the easiest
qualification to meet; we required a minimum of 5000 counts in a given
spectrum, but most of our regions have easily over 10,000, thanks to
the depth of the 2009 observation. Thirdly, we ensured that the 1D
profiles that we extracted, described below, have a constant shape at
the leading edge of the emission. This was done to avoid the situation
that we encountered a few times in Paper II, where the shape of the
emission profile changed between the two epochs. Finally, we attempted
to find knots that were roughly evenly distributed in all sectors of
Tycho. In some places, this was harder than others; the eastern and
southeastern portions of the interior of the remnant were particularly
difficult. The issue was not in finding knots bright enough, but in
finding knots that maintained spatial coherence between the two
epochs.

\begin{figure}[h!tb]
\includegraphics[width=16cm]{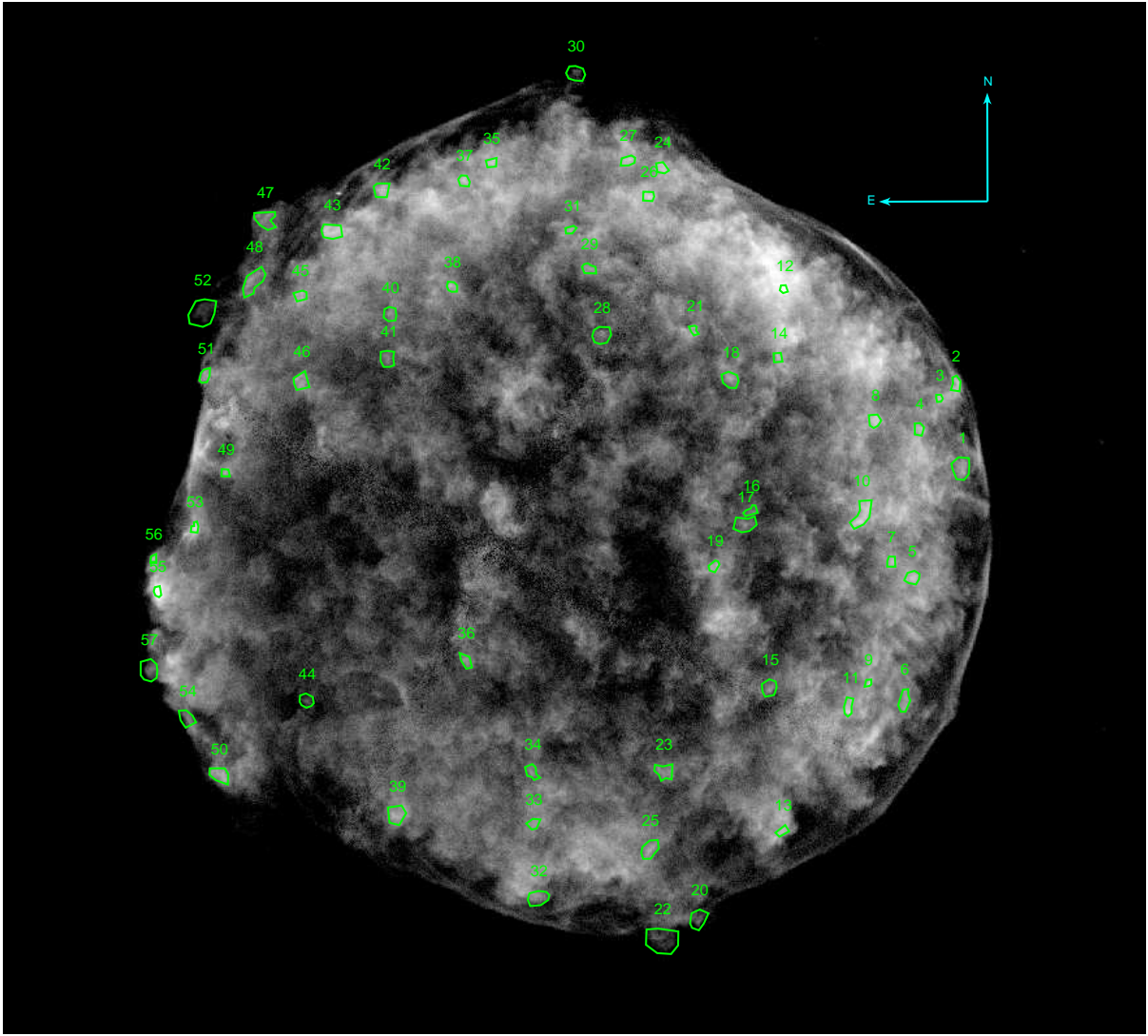}
\caption{A broadband (1-8 keV) {\it Chandra} image, overlaid with the
  57 regions used for the expansion measurements. Tycho's SNR is $\sim
  8.5'$ in diameter.
\label{regions}
}
\end{figure}

We emphasize that it is neither within the scope of this paper nor is
it feasible to account for the motion of every tiny structure within
the remnant. We ended up with a total of 57 ejecta knots that
satisfied all of our conditions, above. These regions are shown in
Figure~\ref{regions}. {\bf For each of the knots,
we drew an extraction region centered on the brightest part of the knot,
with the region cutting off when the flux (as determined from the 
exposure-corrected 2009 flux image) dropped below 1.5 $\times 10^{-7}$
photons cm$^{-2}$ s$^{-1}$.} We used that exact region for the spectral extraction,
using an off-source background from an annulus surrounding the
remnant. We experimented with other choices of off-source background selection,
but found no difference in the resulting spectral fits (the background
accounts for, on average, about 0.5\% of the flux in the 1.2$-$2.8 keV
band that we use for our fitting). {\bf We also experimented with local backgrounds
from within the remnant surrounding each knot. We find, as reported in SH16,
that the local backgrounds vary significantly enough that there is no way to 
know what an accurate local background is. Choosing these local backgrounds
increases the velocities we measure (in both directions: redshifts get redder and
blueshifts get bluer), but in approximately equal amounts of around 1000
km s$^{-1}$. Given the uncertainties of determining the appropriate local
background for each individual knot, we have stayed with the systematic
approach of using a uniform off-source background to be the safest and 
most conservative approach. To measure the proper motion, we
used a ``projection'' region in ds9, drawn along both the X and Y directions.}

\subsection{ACIS-I Gain Calibration}
\label{calibration}

Measuring the velocity along the line of sight is done by measuring
the red and blueshifts of the Si and S lines that dominate the
spectra. We detail in Section~\ref{measurements} how this was done,
and detail the statistical uncertainties of these measurements. We
discuss here the systematic uncertainty introduced by the accuracy of
{\it Chandra's} gain calibration. To better understand this, we
consulted with {\it Chandra} calibration experts, who assisted us in
quantifying this (P. Plucinsky and N. Durham, private communication).

The gain for the ACIS-I instrument is calibrated using three onboard
sources with spectral lines at known energies: an Al K$\alpha$ line at
1.487 keV, a Ti K$\alpha$ line at 4.511 keV, and a Mn K$\alpha$ line
at 5.898 keV. The {\it Chandra} calibration team went back to these
2009 observations and compared the measured line energies for all
three of these lines for the entire ACIS-I array. The focal plane
temperature (which can affect the gain) was stable during these
observations. They found that for virtually the entire ACIS-I array,
the gain calibration is good to a level of $<$0.3\%, with some places
being significantly better, particularly for the Ti and Mn K$\alpha$
lines. However, because our measured lines of Si and S lie closer to
the Al K$\alpha$ calibration line, we adopt the conservative
systematic uncertainty of 0.3\%, or 900 km s$^{-1}$ for our line of
sight (V$_{Z}$) velocities.

\begin{figure}[h!tb]
\includegraphics[width=16cm]{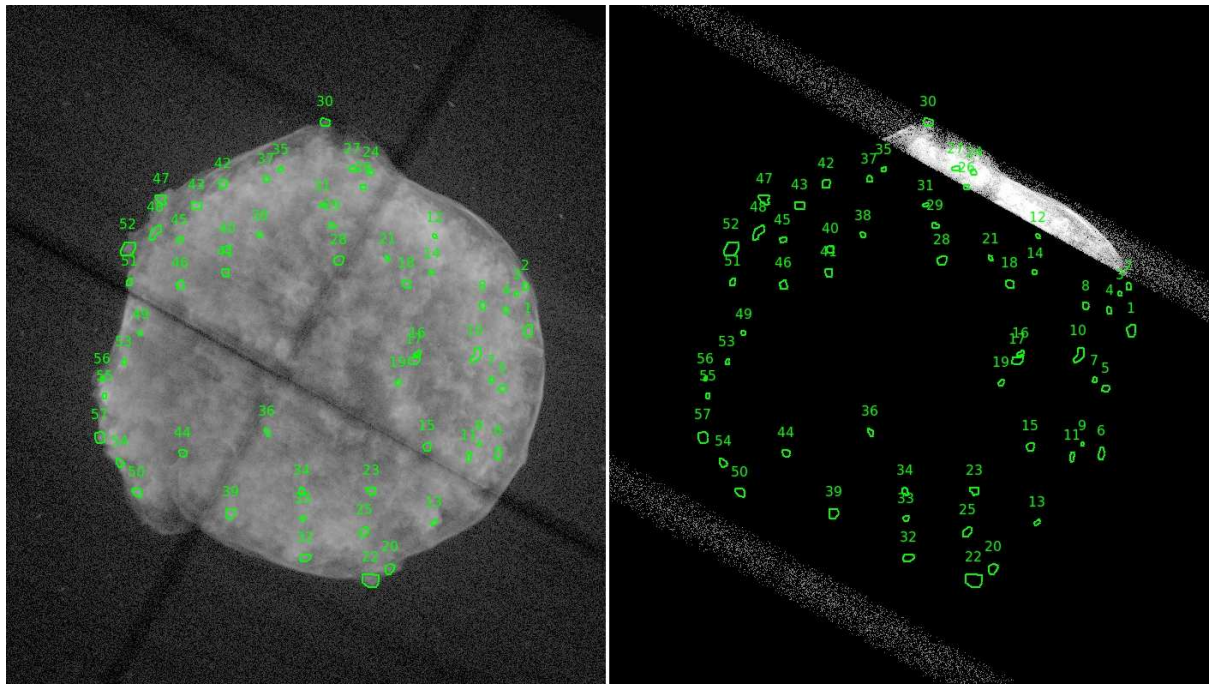}
\caption{{\it Left}: the events file for the merged 2009 ACIS-I
  observation, overplotted with our measurement regions, shown for
  reference. {\it Right}: The events file with only the chip rows that
  fall outside of a calibration range of 0.3\%, overplotted with the
  same regions.
\label{badchips}
}
\end{figure}

The exceptions to this are the center rows of all 4 ACIS-I chips,
where the calibration is worse for the Al K$\alpha$ line in
particular. In some places, the measured values of the calibration
lines differ by up to 1\% from the fiducial values. The 2009
observation, like most {\it Chandra} observations of Tycho, used the
ACIS-I array, centered on the approximate center of the
remnant. Because of this, most of the remnant {\em does not} fall on
the centers of the four chips. To check whether any of our regions
fell in the ``bad'' calibration regions, we generated an image of the
remnant containing only the center rows in which the calibration
uncertainty was greater than 0.3\%. We then overplotted our regions,
and found that only four of them (regions 24, 26, 27, and 30) lie
within the affected rows. This is shown in Figure~\ref{badchips}. We
still report measured values for those four regions, but we note that
the uncertainty could be as high as 1\%, or 3000 km s$^{-1}$.

\section{Measurements}
\label{measurements}

\subsection{Proper Motions}

\begin{figure}[h!]
\includegraphics[width=5.5cm]{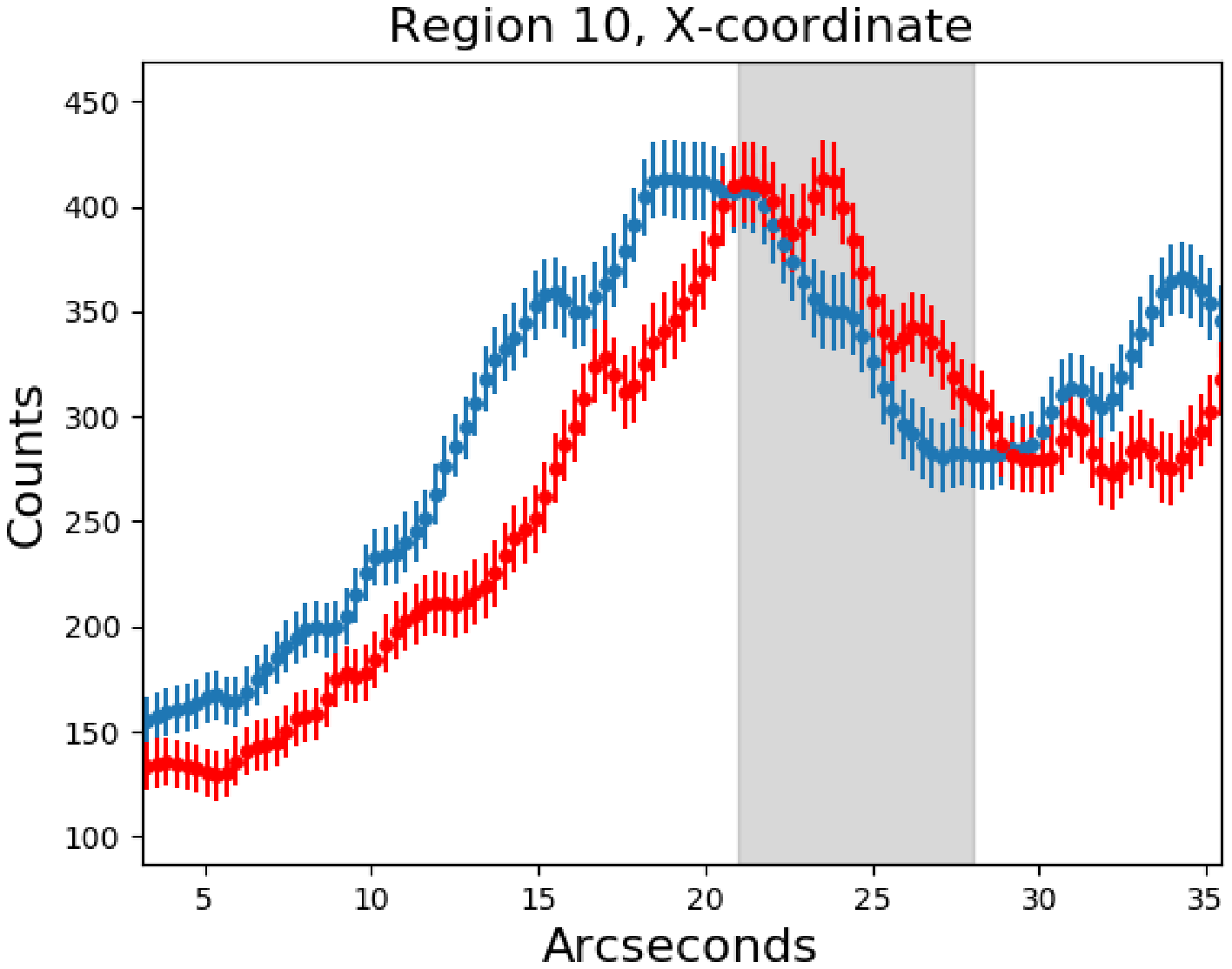}
\includegraphics[width=5.5cm]{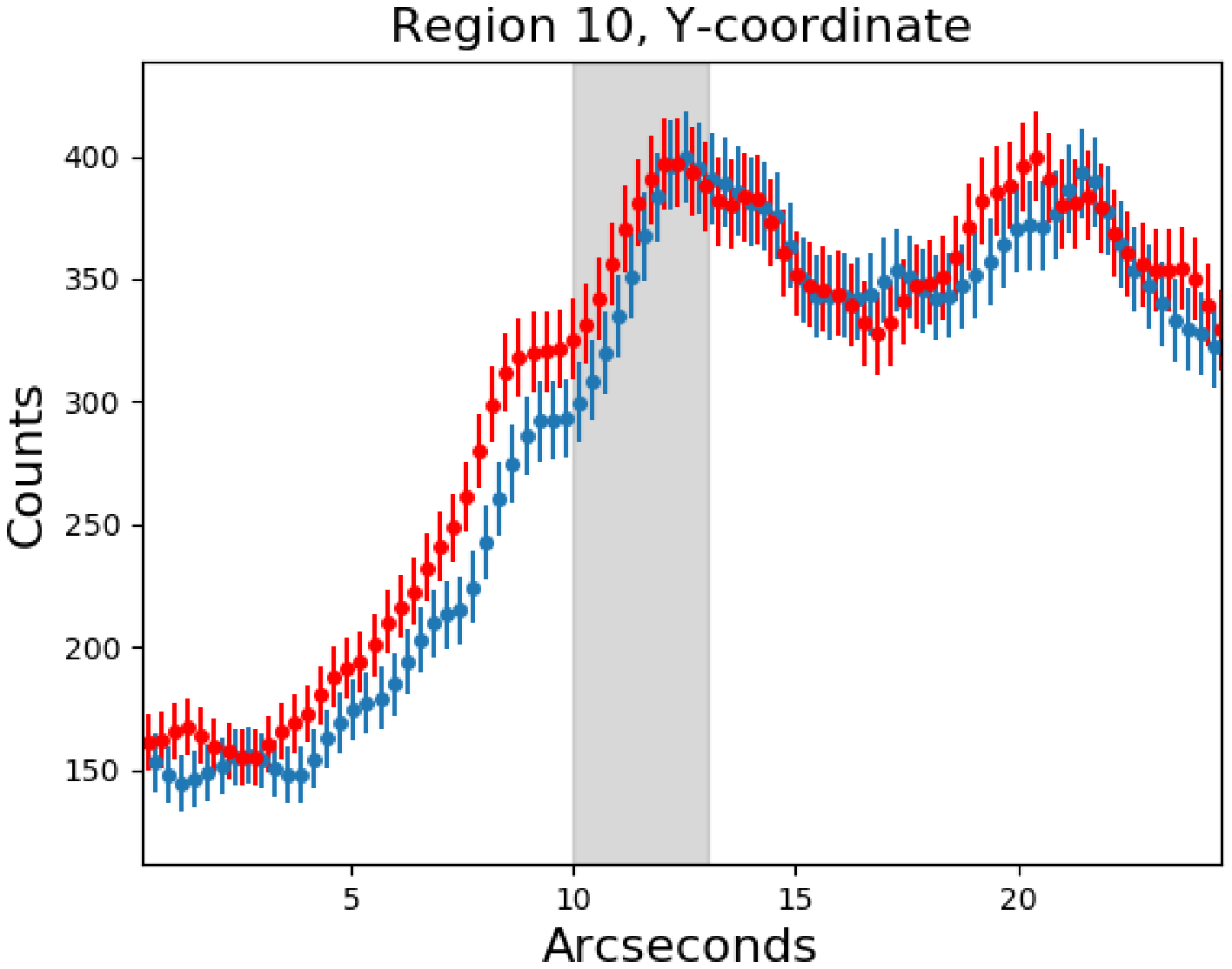}
\includegraphics[width=5.5cm]{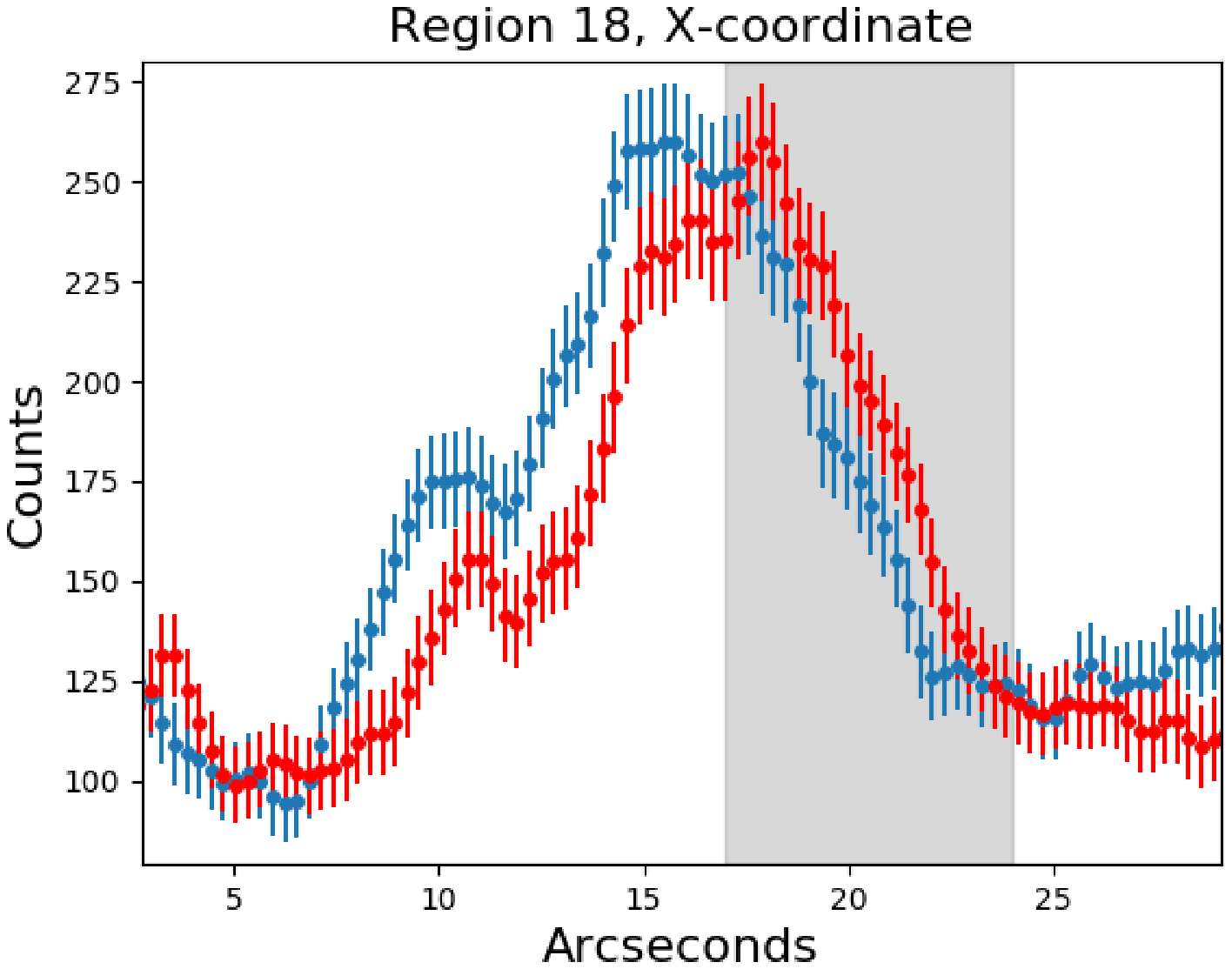}
\includegraphics[width=5.5cm]{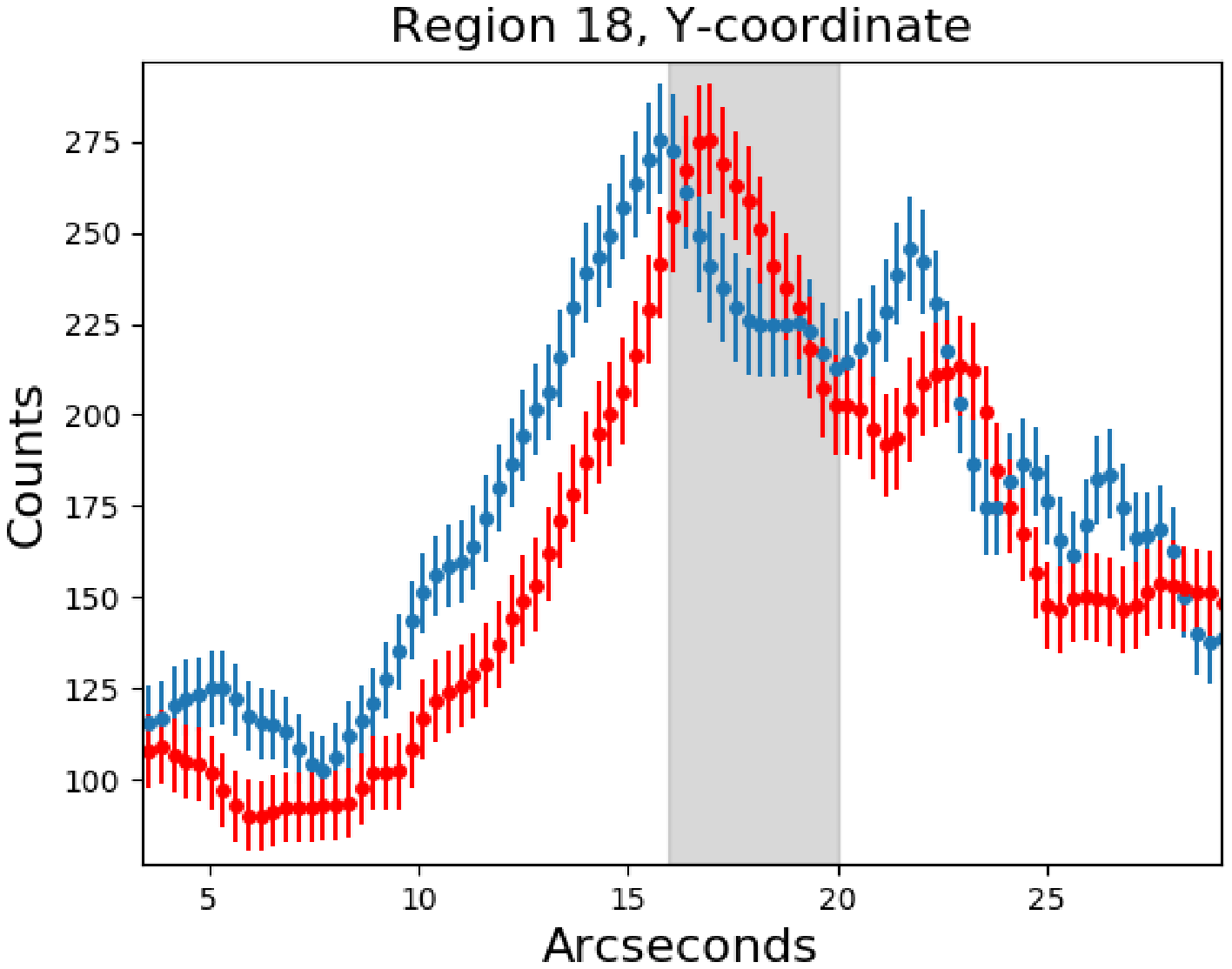}
\includegraphics[width=5.5cm]{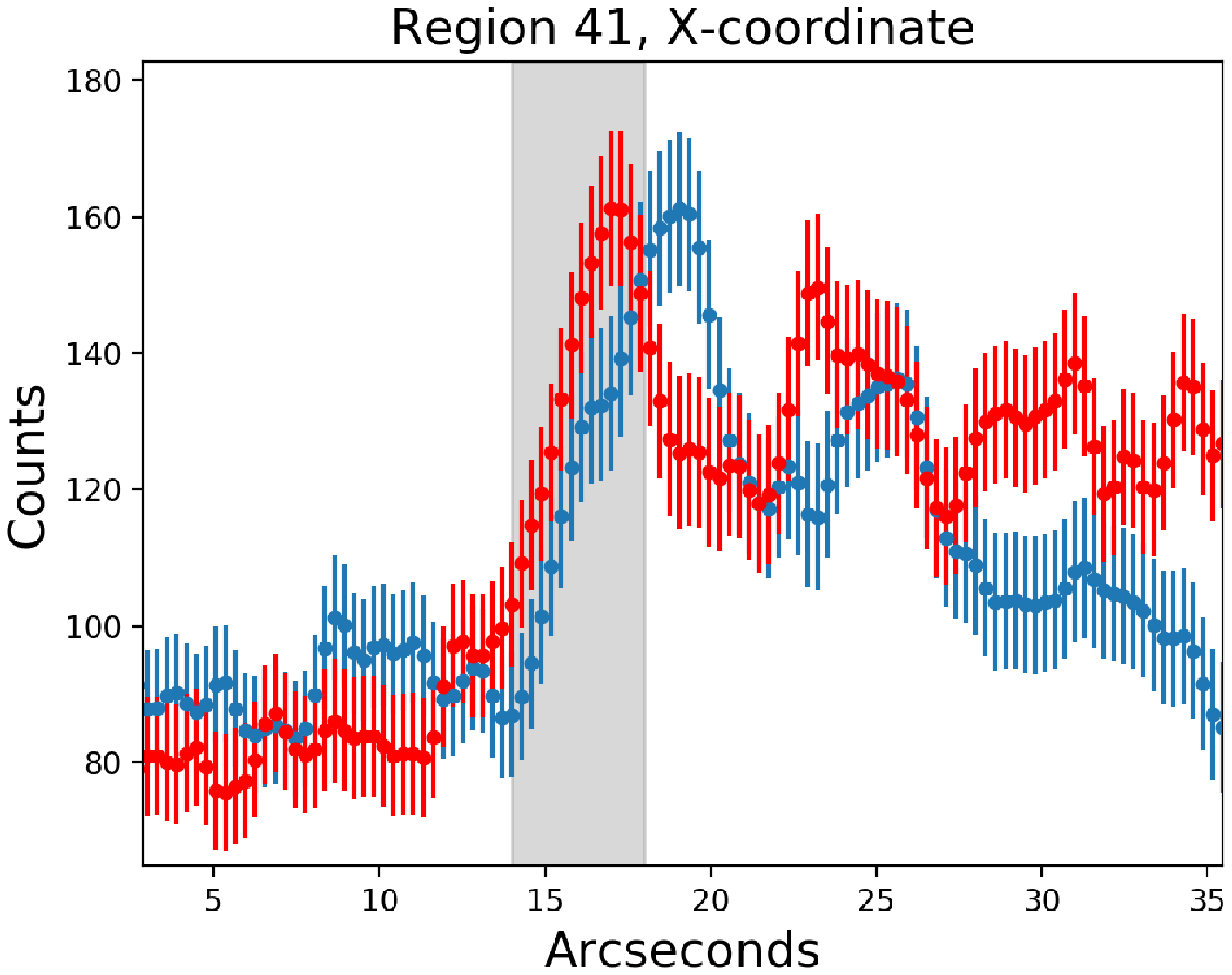}
\includegraphics[width=5.5cm]{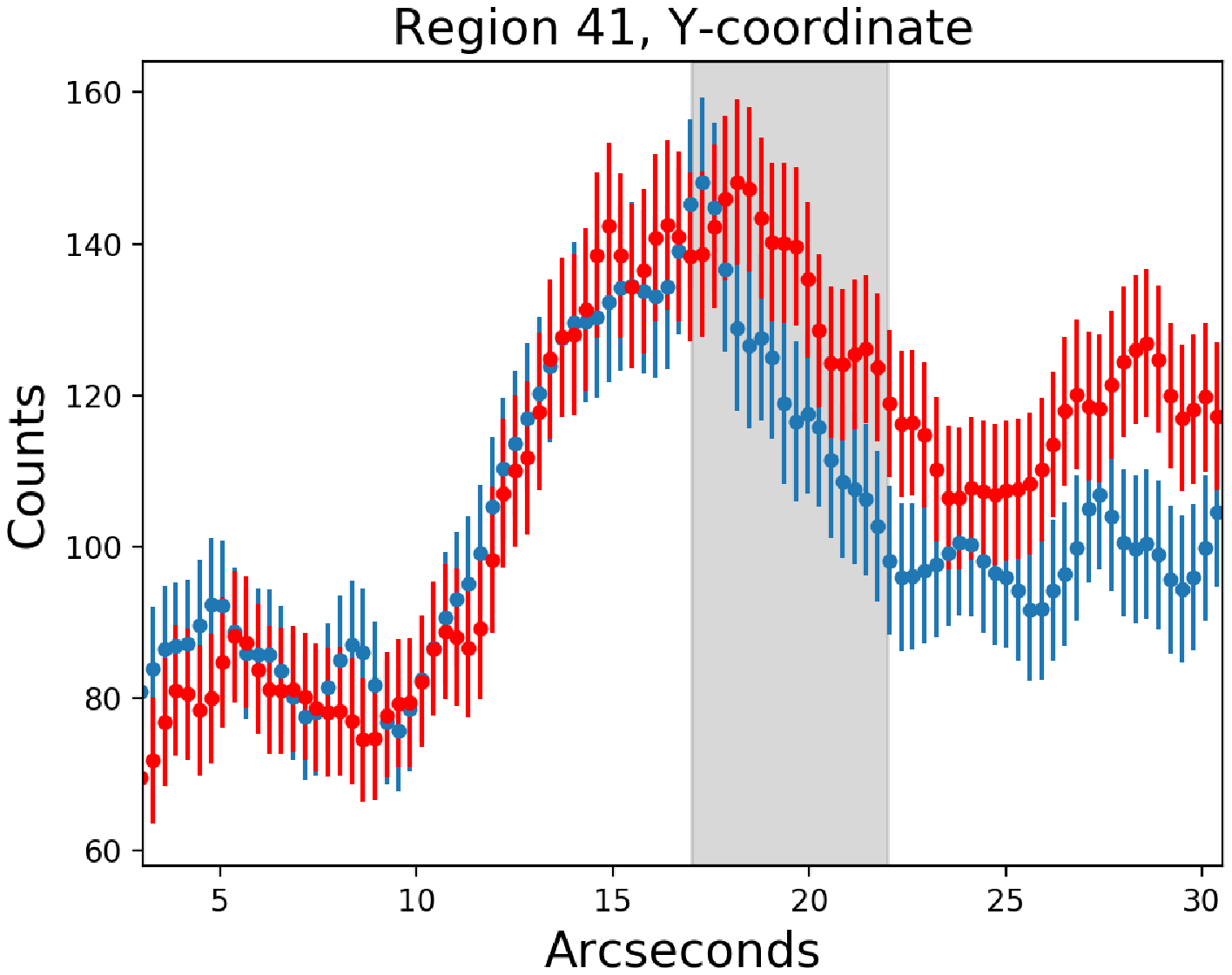}
\includegraphics[width=5.5cm]{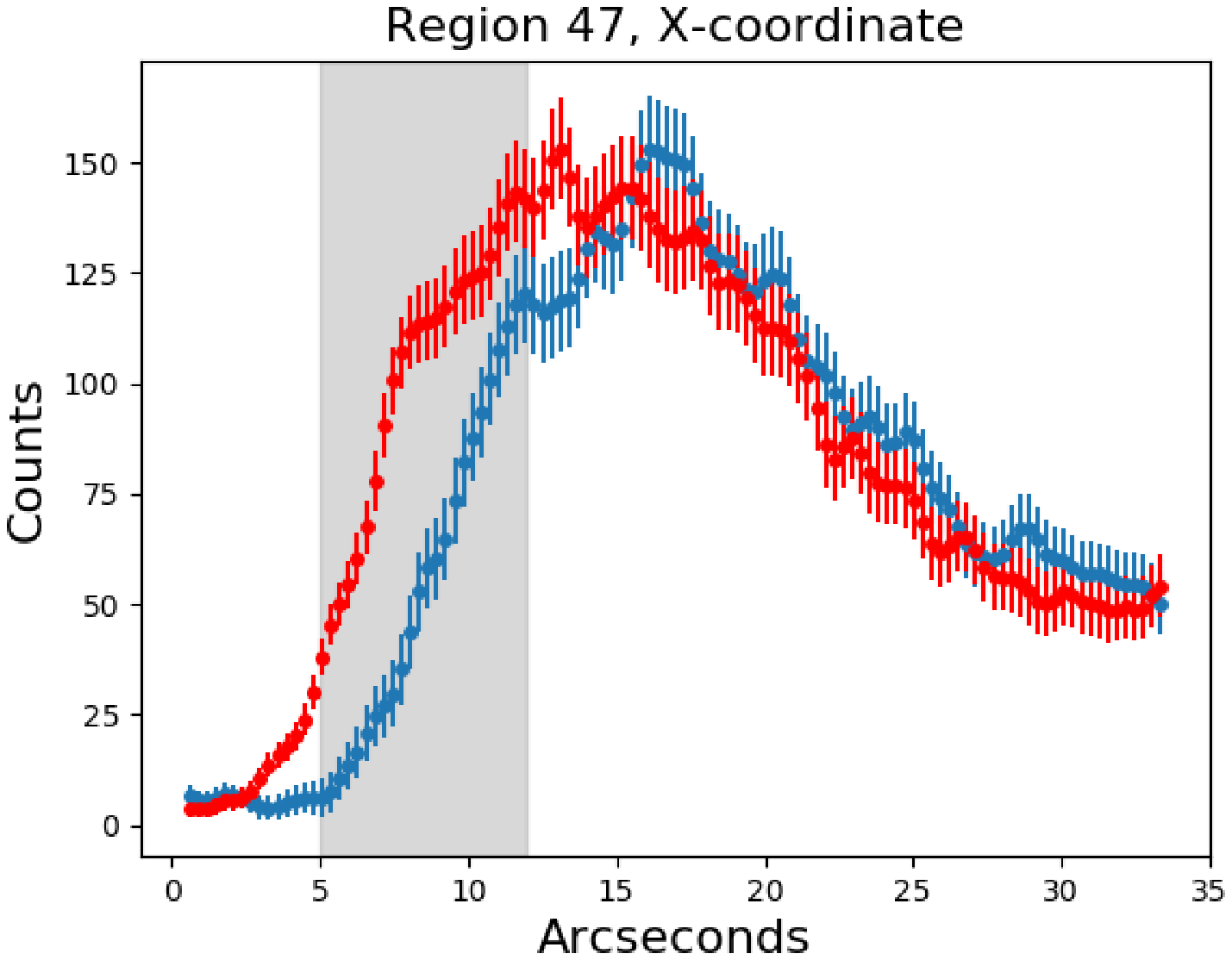}
\includegraphics[width=5.5cm]{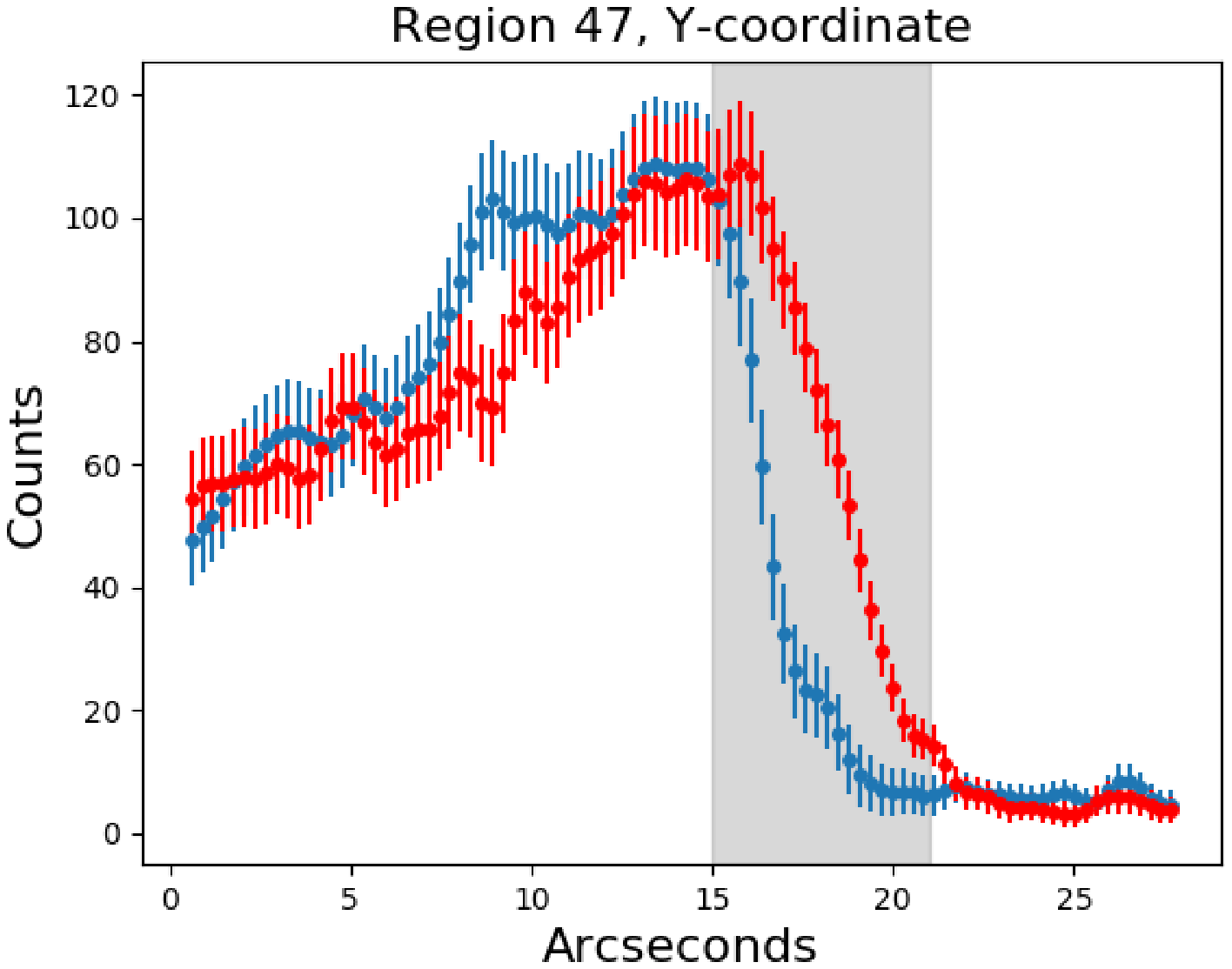}
\caption{An example of our 1D profiles from several sample regions. We fit 
the leading edge of the emission, shifting epoch 1 (blue profile) to fit epoch
2 (red) in the region containing the leading edge of the emission in epoch 2 
(grey region).
\label{profiles}
}
\end{figure}

Our procedure for measuring the proper motions follows that of Paper
II, which is based on \citet{katsuda08} and used in other SNR works,
such as \citet{winkler14} and \citet{yamaguchi16}. We extract the 1D
radial profiles from both epochs in both the X and Y (R.A. and Dec.)
directions, with uncertainties (where the uncertainty on each pixel is
the square root of the number of counts in that pixel, which is summed
across the width of the projection region), and shift epoch 1 relative
to epoch 2. We extract the profiles in pixel space, with shifts
calculated on a grid of 2000 points with a size of 0.025 pixels. A fit
is obtained when the value of $\chi^{2}$ is minimized, and the 90\%
uncertainties we report in Table~\ref{velocities} come from a shift in
$\chi^{2}$ of 2.71 in either direction. As in Paper II, we fit
for the shift in an area containing the leading edge of the filament. {\bf We show
the X and Y projections from several sample regions in Figure~\ref{profiles}. 
Regions that maintain their shape particularly well have very small errors; in some 
cases, less than 10 km s$^{-1}$.}

\begin{figure}[h!tb]
\centering
\includegraphics[width=12cm]{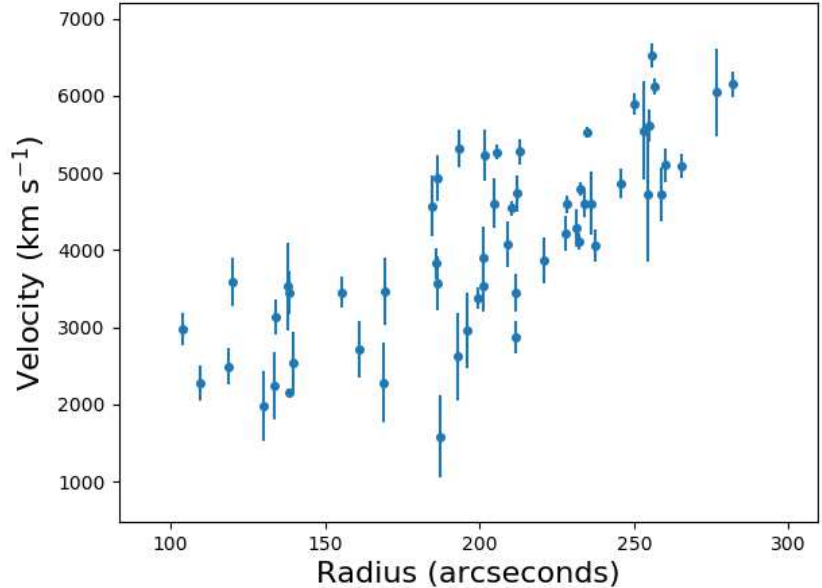}
\caption{Total velocity in the plane of the sky plotted against the radius from
the explosion site, as described in the text. The radius depends slightly on the choice
of explosion site (see description in the text), though the velocity is independent of this
and only depends on the measured proper motions, known time baseline, and the 
assumed distance of 3.5 kpc.
\label{vrad}
}
\end{figure}

Converting a measured proper motion to an absolute velocity requires
knowing the distance to the remnant. The reported value of the
distance to Tycho has varied in the literature over time. An analysis
by \citet{chevalier80} suggested a distance of 2.3 kpc, a similar
distance to that reported in \citet{albinson86}. Later work by
\citet{schwarz95} favored a distance of over 4 kpc. In Paper I, we
compared hydrodynamic simulations to the observations, preferring a
distance of 3.5 kpc, a distance that fits nicely with the 3.8 kpc
recently derived by SH16. For the purposes of this paper, we adopt a
distance of 3.5 kpc, and scale proper motion velocities
accordingly, noting that the results of this paper are not dependent on knowing the absolute distance. 
The proper motions for our 57 regions are listed in
Table~\ref{velocities}. As expected, higher values for the proper
motion are found in the regions along the edge of the remnant, while
lower ones are found for those closer to the interior. In several
regions where the ejecta are clearly seen in the image to be
protruding in front of the nonthermal rim (e.g., regions 22, 30, and
57), we measure proper motion velocities in excess of 6000 km
s$^{-1}$. {\bf In Figure~\ref{vrad}, we show the radial velocities in the plane of the 
sky (i.e., the combined X and Y velocities) as a function of their radial distance
from the presumed explosion site that we found in Paper II ($\alpha$ =
0$^{h}$25$^{m}$22.6$^{s}$ and $\delta$ = 64$^{\circ}$8$'$32.7$''$). 

This choice of explosion site is offset by about 23$''$ from the geometric center of the
remnant. The choice of this explosion site versus the geometric center of \citet{ruiz04} 
has little effect on the radii and deceleration parameters measured. 
Along regions located in the SE and NW portions, the effect is nearly zero, while the
most affected region, region 19, differs by about 20\%. The average effect is $\sim 9$\%. }

\subsection{Doppler Velocities}

While the proper motion measurements can give us a velocity in the
plane of the sky (modulo the distance to the remnant), the LOS
velocity is needed for a total spatial velocity of any particular
ejecta knot. SH16 looked at the doppler shifts of several ``blobs'' of
ejecta in Tycho, mostly in the center of the remnant, and converted
these doppler shifts to a spatial velocity. We follow a similar
procedure here, applying it to our regions, which are outside the
center of the remnant, and thus can be combined with the measured
proper motions. 

The ejecta emission from Tycho is dominated by remarkably strong
emission from the Si XIII K$\alpha$ line at around 1.85 keV, a line
commonly seen in young Type Ia SNRs. Another strong line from S XV is
present at about 2.45 keV, and weaker lines from Mg, Ar, Ca, and Fe
are also present in the spectrum. We focus here on the energy range
from $1.2-2.8$ keV, where the Mg, Si, and S lines appear. We use two
independent procedures to determine the red/blueshift of each of our
57 regions using this energy range. For one method, used in SH16, we
fit a non-equilibrium ionization (NEI) model to the spectrum, modeling
the temperature and ionization state of the plasma and fitting an
overall red/blueshift to the entire spectrum. As an alternative check,
we fit Gaussian line profiles to each of the lines in the spectrum,
fitting the centroid, and thus the doppler shift, of {\it only} the Si
K$\alpha$ line.

In modeling the spectrum with a thermal model, we use an absorbed NEI
model with an underlying power-law component to account for the
possibility of any non-thermal emission that might be present along
the line of sight. We fix the absorbing column density at $6 \times
10^{21}$ cm$^{-2}$, consistent with the values reported in
\citet{hwang02} and SH16, but we note that the exact value of this
parameter (within the reasonable range as reported in the literature
of $5-10 \times 10^{21}$ cm$^{-2}$) has no effect on the fits. We also
fix the value of the power-law index to 2.6, consistent with the
values found in \citet{cassamchenai07} and \citet{tran15}. This value
also has no effect on the fits, as in almost all of our 57 cases, the
normalization of the power-law component simply drops to zero, as
expected for emission in the interior of the remnant. Nonetheless, we
keep this component in for completeness. In the NEI model, the
temperature and ionization timescale ($\tau_{i} \equiv \int^{t}_{0}
n_{e} dt$) are allowed to vary freely, as are the abundances of Mg,
Si, and S. We fit for the doppler shift of the NEI model, as a whole,
and convert this to a velocity. The uncertainties we report are the
90\% confidence intervals for the value of the redshift. We list the
fit parameters of the NEI model in Table~\ref{neifits}.

As a check on these numbers, we apply a different model, consisting of
Gaussian lines for the Mg, Si, and S lines, on top of an absorbed
thermal bremsstrahlung model with an underlying power-law
component. The absorbing column density and power-law index are fixed
to those above, while the temperature of the thermal continuum is
fixed to 1 keV. In order to translate a measured line centroid into a
doppler velocity shift, we first need to determine the rest energy of
the Si K$\alpha$ line, which has a dependency on temperature and
ionization state. The emission from Si in Tycho is somewhat
complicated and consists of emission from both H and He-like
states. The dominant line around 1.85 keV is actually a triplet,
consisting of a resonance line with a rest energy of 1.865 keV, an
intercombination line, with a rest energy of 1.854 keV (actually, this
line itself is a doublet, but the lines are so close as to essentially
be one line), and a forbidden line, with a rest energy of 1.840 keV.

These lines are not resolvable with CCD spectrometers such as those on
{\it Chandra}, and thus blend into one line, which we refer to here as
the Si K$\alpha$ line. Additionally, two other lines from the K$\beta$
and K$\gamma$ transitions occur at 2.183 and 2.294 keV,
respectively. Finally, a Ly$\alpha$ line from Si XIV appears at 2.006
keV, which blends with the Si K$\alpha$ line. While it cannot be
resolved from the K$\alpha$ line, the Ly$\alpha$ line can be seen as
an asymmetric ``tail'' on the blue side of the K$\alpha$ line. 

Our model for these lines consists of seven Gaussian components on top
of the thermal bremsstrahlung and power-law components. The seven
components correspond to the Si K$\alpha$, K$\beta$, K$\gamma$, and
Ly$\alpha$ lines, as well as one component for the Mg K$\alpha$ line
at $\sim 1.35$ keV, and two more for the K$\alpha$ and Ly$\alpha$
lines of S at 2.45 and 2.62 keV, respectively. In general, the line
centroids are allowed to float freely, with the caveats that the Si
K$\gamma$ line is fixed to a centroid that is 0.111 keV higher than
the K$\beta$ line, and the line widths for all Si species are tied
together (as are those for S).

From the theoretical side, we calculated the centroid of the Si
K$\alpha$ line on a two-dimensional grid in temperature and ionization
timescale parameter space. From our NEI model fits to each region, we
find that the temperature ranges from 0.57 to 2.61 keV, while the
ionization timescale varies from $\sim 2 \times 10^{10} - 1 \times
10^{12}$ cm$^{-3}$ s, though most values fall in a much smaller range
of $4 \times 10^{10} - 2 \times 10^{11}$ cm$^{-3}$ s. Nonetheless, if
we take the most extreme values for both quantities, we find that the
Si K$\alpha$ centroid varies over only a small range, from 1.8558 to
1.8582 keV, with an average value of 1.8570 keV. We use this value as
the ``fixed'' rest energy, noting that the systematic uncertainties in
either direction are 1.2 eV, or 190 km s$^{-1}$, significantly smaller
than the statistical errors on the NEI fits, above.

The Si Ly$\alpha$ line at 2.006 keV is not affected by temperature or
ionization state, so its shift will {\em only} be affected by the
doppler shift resulting from the LOS velocity of each ejecta knot. In
principle, an accurate determination of the Ly$\alpha$ line centroid
could solely be used to measure the doppler velocity, but this would
require an X-ray micro-calorimeter with high spectral and spatial
resolution. While we cannot, using CCD spectra, accurately determine
the Ly$\alpha$ centroid, we do fix it in the model to a value of 0.149
keV (2.006 - 1.857 keV) higher than that of Si K$\alpha$, forcing
these two lines to move together. We fit the same $1.2-2.8$ keV energy
range as before. The fits are statistically better for this model, but
this is expected, given the increased number of free parameters within
the model.

\begin{figure*}[h!tb]
\includegraphics[width=8cm]{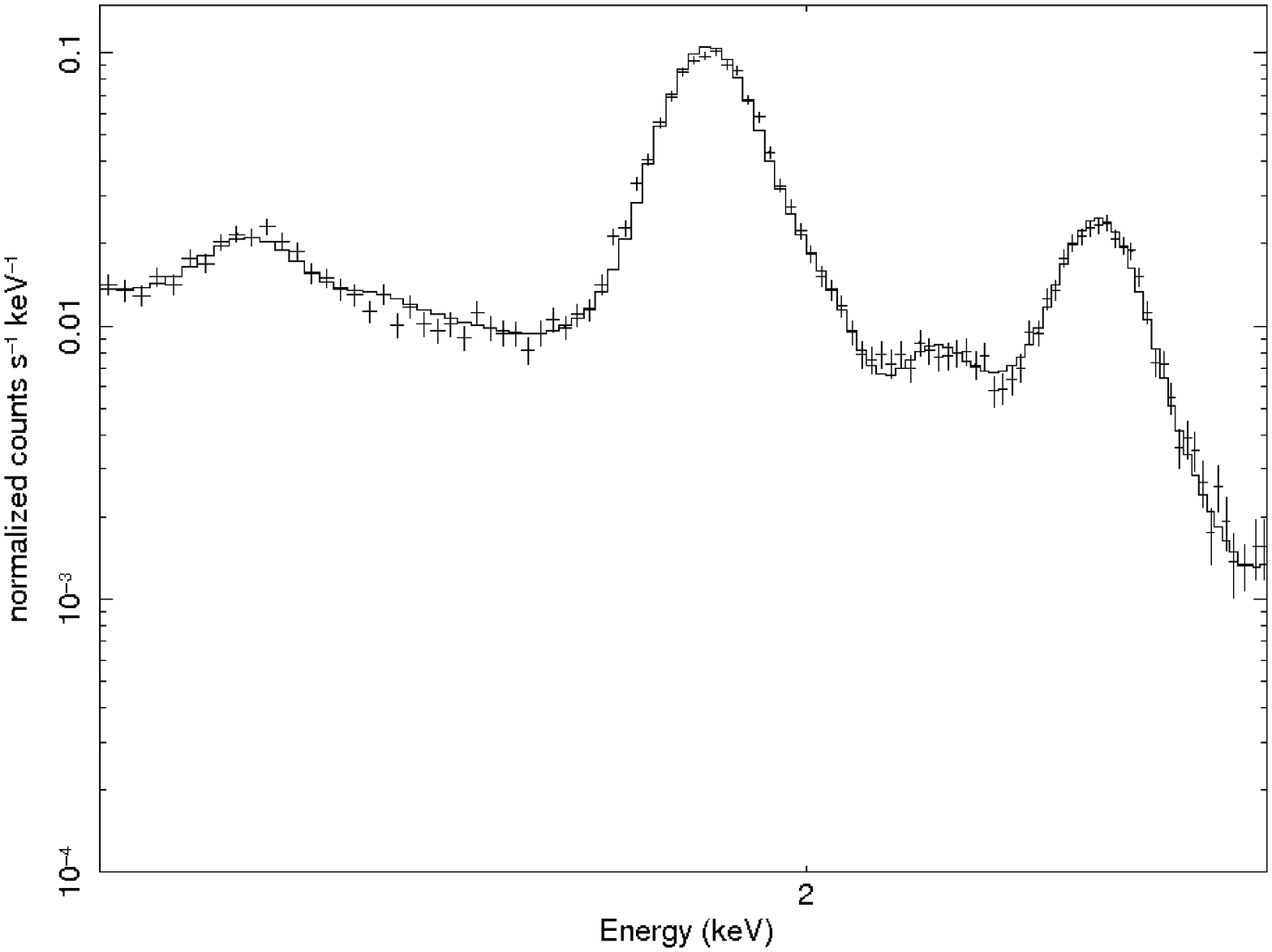}
\includegraphics[width=8cm]{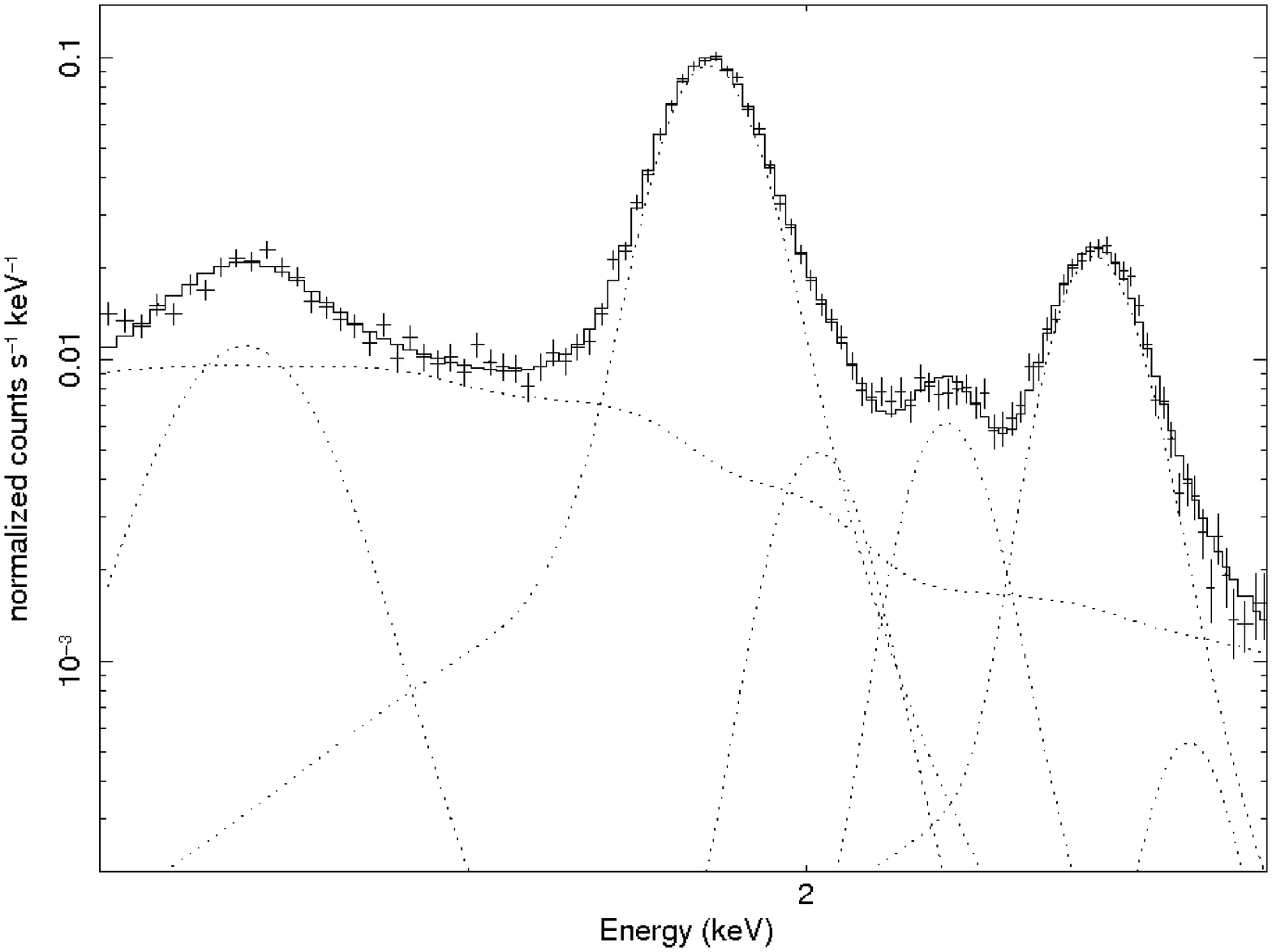}
\caption{{\it Left:} The $1.2-2.8$ keV spectrum from region 46, fit
  with the NEI thermal model described in the text. {\it Right}: The
  same data, but fit with the Gaussian line model described in the
  text. The continuum is a thermal bremmstrahlung model with
  temperature fixed to 1 keV. From left to right, the lines are Mg
  K$\alpha$, Si K$\alpha$, Si Ly$\alpha$, Si K$\beta$, S K$\alpha$,
  and S Ly$\alpha$.
\label{modelfits}
}
\end{figure*}

\begin{figure}[h!tb]
\centering
\includegraphics[width=15cm]{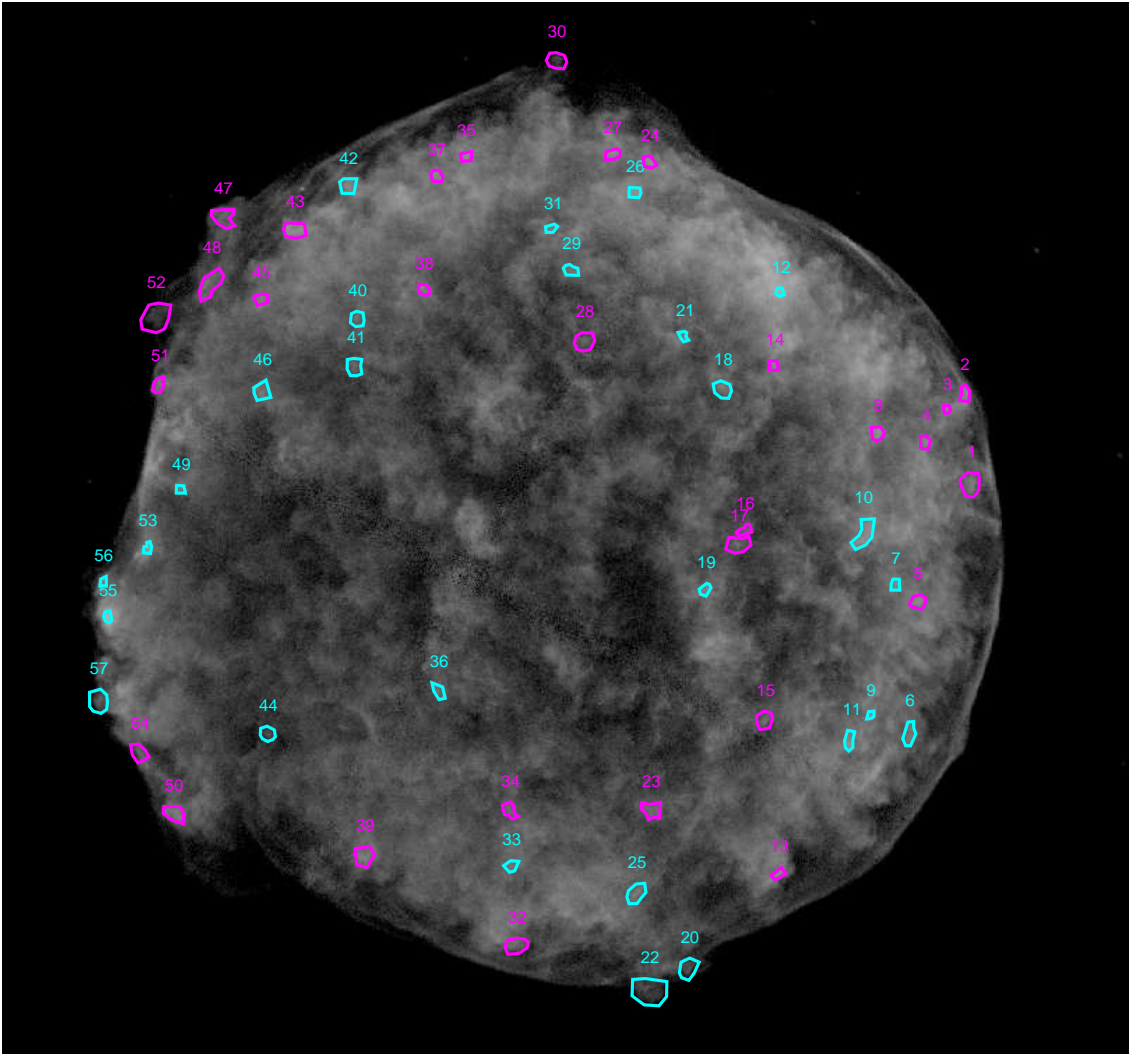}
\caption{This is identical to Figure~\ref{regions}, except that the regions have
been color-coded magenta and cyan to indicate whether the knot is redshifted or
blueshifted, respectively.
\label{red_blue}
}
\end{figure}

In 52 out of our 57 regions, the value for the red/blueshift of the Si
K$\alpha$ line using this Gaussian line method agrees within errors of
that measured from the NEI model fit. Even in the other five, the
values are not significantly different, but fall just outside of the
range of uncertainties. That these two different methods agree so well
increases our confidence in the robustness of our NEI model fits, and
we report the velocities measured from that model in
Table~\ref{velocities}. We show an example of both an NEI model fit
and a Gaussian line model fit to one of our regions, region 46, in
Figure~\ref{modelfits}. As expected, the doppler velocities are lower
than those reported in SH16, since their numbers come mostly from
ejecta knots near the center of the remnant, while ours come from the
outer portions. We find a wide range of both redshifted and
blueshifted velocities, ranging from 320 to 3360 km s$^{-1}$. Even
though our regions were chosen blindly, with no {\it a priori}
knowledge of the doppler shift in a given ejecta knot, we find roughly
equal numbers of red and blueshifted knots (30 red, 27 blue). {\bf As shown
in Figure~\ref{red_blue}, the redshifted and blueshifted knots are roughly 
evenly distributed throughout the remnant, and in some cases, knots in relatively 
close spatial proximity are moving in different directions. SH16 show a similar
figure in their work, except that their regions are in the center of the remnant.} 

\subsection{Total Velocities}

With the proper motion and LOS velocities calculated for all 57 of our
regions, we can construct total velocity vectors for each where both
the magnitude and direction are known. We report the total velocities
and their components in Table~\ref{velocities}. For the velocity
components, we use a Cartesian coordinate system, where the X-Y plane
corresponds to the plane of the sky, with X-positive to the right
(west) and Y-positive in the up (north) direction. The Z-axis is
positive moving away from the observer (redshift) and negative towards
the observer (blueshift). We find total velocities that range from
just under 2400 km s$^{-1}$ to nearly 6600 km s$^{-1}$, with mean and
median values of 4430 and 4450 km s$^{-1}$, respectively. {\bf This is quite close
to the value of 4700 km s$^{-1}$ found by \citet{hayato10}.}

To construct a velocity vector for each knot for display purposes, one
must know the position of the knot within the remnant. While the X and
Y positions simply correspond to the R.A. and Dec. of the knots'
coordinates, there is no inherent way of knowing the
Z-coordinate. Still, we can make some reasonable assumptions. The
knots should move radially from the explosion site out, and we assume
that Tycho is spherically symmetric with a radius of 4.33 pc (4.2' at
3.5 kpc). The latter is reasonably well justified by noting the
near-perfect circular shape of the remnant on they sky (while it is
possible that the remnant has the shape of, e.g., a prolate spheroid
viewed ``end-on,'' this requires our position to be coincidentally
located along the major axis, which is unlikely), as well as by the
fact that our proper motion velocities at the extreme edge of the
remnant are quite well matched to the doppler shifted velocities of
the knots in the center of the remnant as measured by SH16. We
therefore assume that the Z position is simply the Z-velocity
multiplied by the age of the remnant (437 yr in the 2009 epoch),
normalized to the highest proper motion value we measure (6517 km
s$^{-1}$. For example, whatever its X and Y coordinates, a knot with a
redshift of 3260 km s$^{-1}$ would be halfway between the center of
the remnant and the far edge of the shell.

\section{Discussion}



\begin{figure}[h!tb]
\includegraphics[width=7cm]{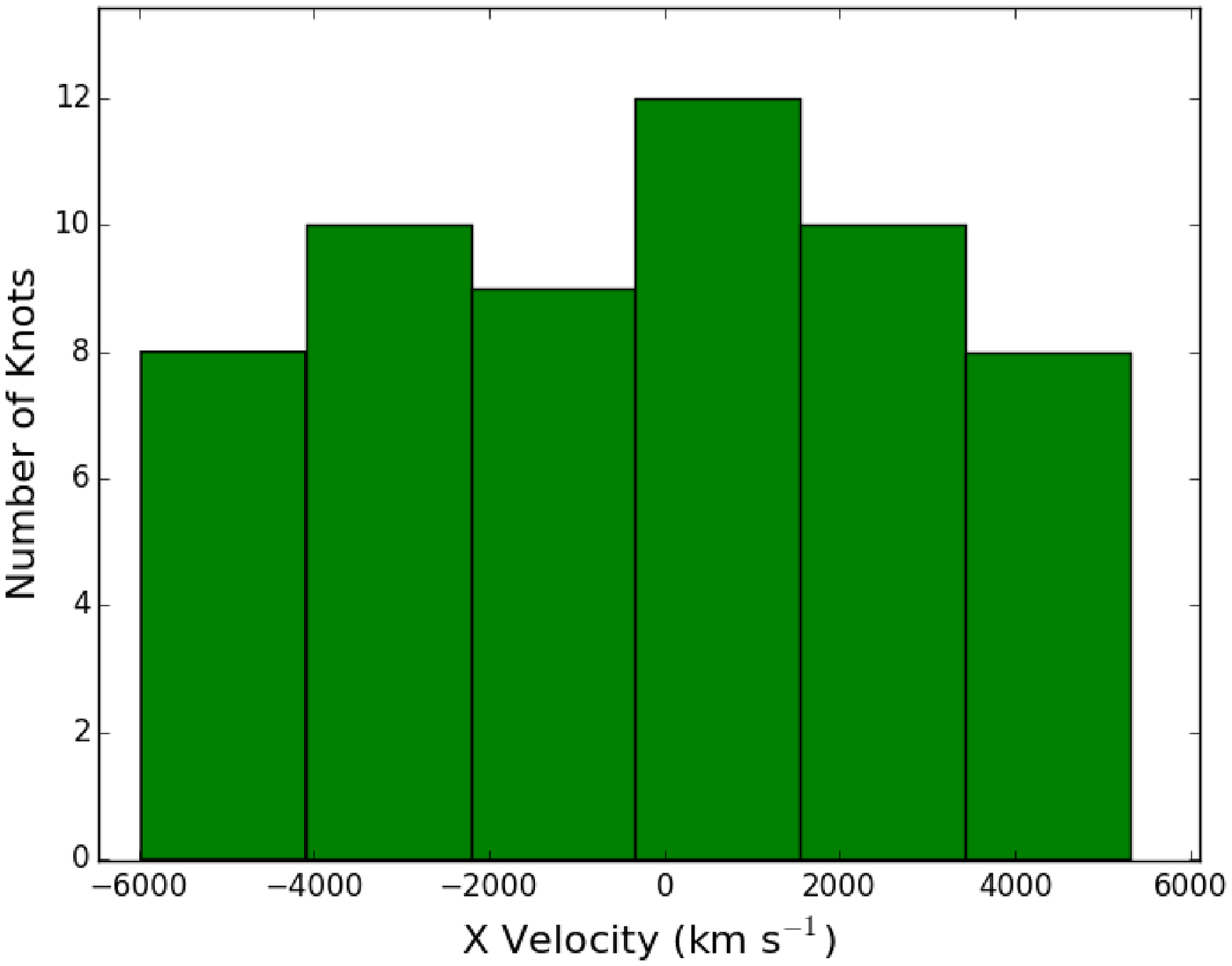}
\includegraphics[width=7cm]{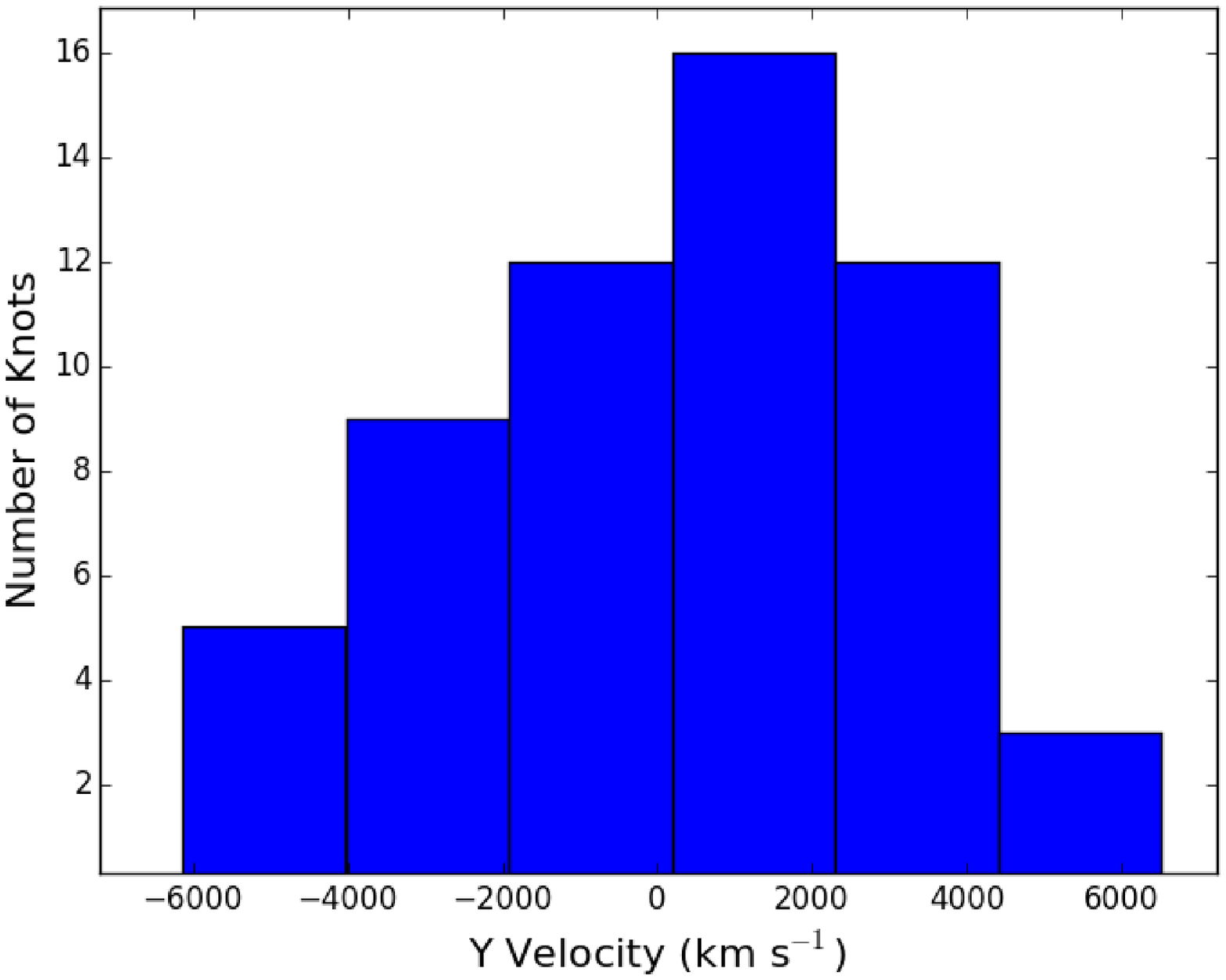}
\includegraphics[width=7cm]{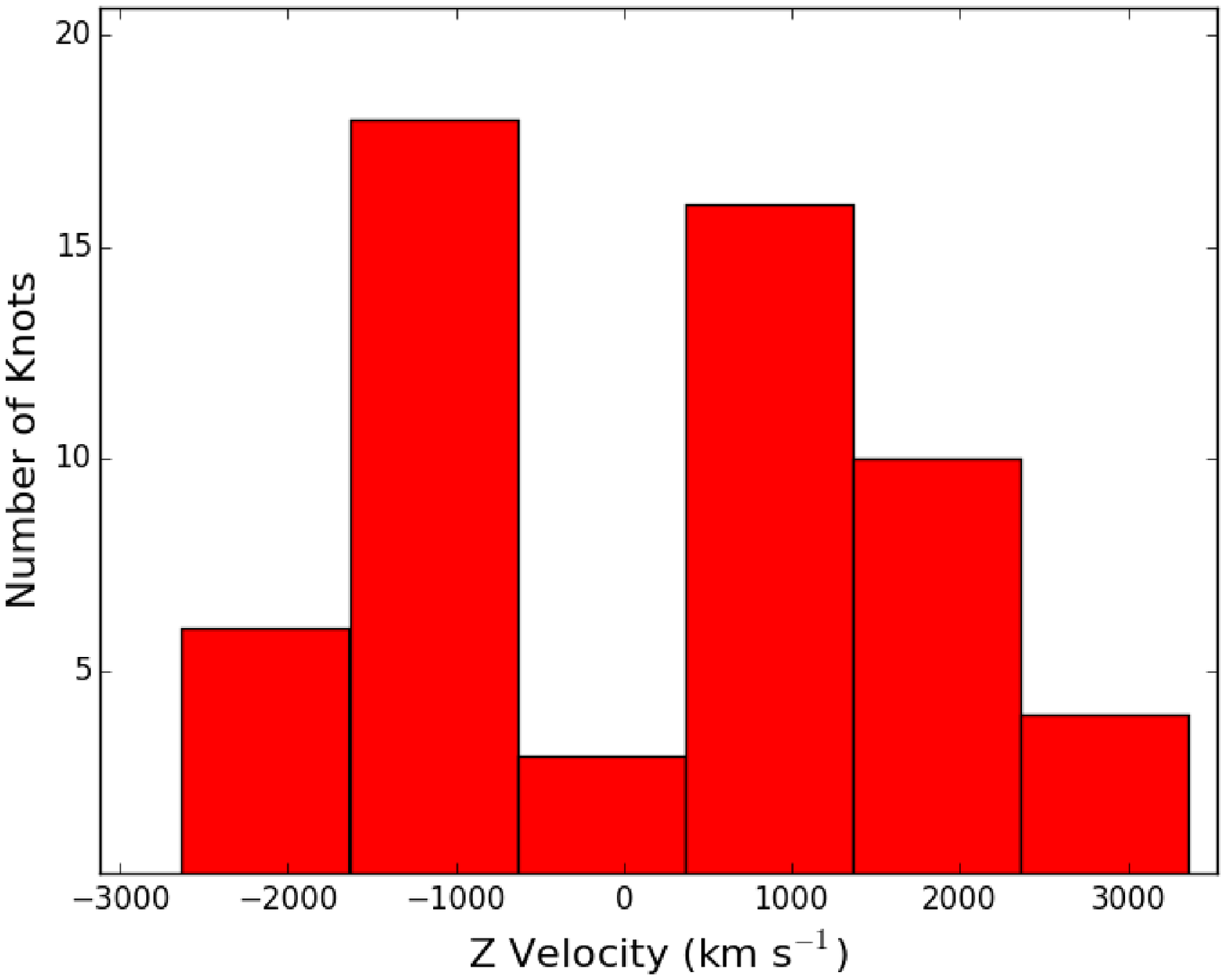}
\caption{Histograms of the velocity distribution in the X (green), Y
  (blue) and Z (red) directions.
\label{histograms}
}
\end{figure}

Quantitatively speaking, we can average the velocities in the X, Y,
and Z directions. We account for the selection effect of the location
of the blobs by weighting the averages to account for the number of
blobs present in, for example, the positive and negative X-coordinate
(the right and left halves of the remnant). We find average velocities
of -179, -88, and 83 km s$^{-1}$ in the X, Y, and Z directions,
respectively. These are negligible compared to the magnitudes of the
velocity vectors themselves. We can further quantify this by making
histograms of the velocity distribution of the knots in each
direction. We show these in Figure~\ref{histograms}. The limited
number of knots forces us to use relatively large velocity bins of
$\sim 2000$ km s$^{-1}$ for these plots; nonetheless, the
distributions around zero are fairly symmetric even with such coarse
bins. The Z-distribution has a minimum around zero because the only
knots that would have exactly zero velocity in the Z-direction are
those along the outermost periphery of the remnant (i.e., moving
entirely in the plane of the sky), and we have relatively few of
those. We see no evidence for an asymmetric expansion of the ejecta in
Tycho.

It is interesting to note the difference between the expansion of the
ejecta, which appears symmetric, and the expansion of the forward
shock wave, delineated by synchrotron emission surrounding the
periphery of Tycho. In Paper II, we show that the forward shock is
expanding significantly faster in the western hemiphere than in the
east (and particularly northeast). We interpreted this, along with our
results from Paper I, as evidence for a density gradient in the ISM
that has begun to slow the blast wave in the eastern portions of the
remnant. Likely due to their high density contrast, the clumps of
ejecta that we measure here have not yet been significantly
decelerated by the more dense material on the east and northeast sides
of the remnant.

\begin{figure}[h!tb]
\centering
\includegraphics[width=9cm]{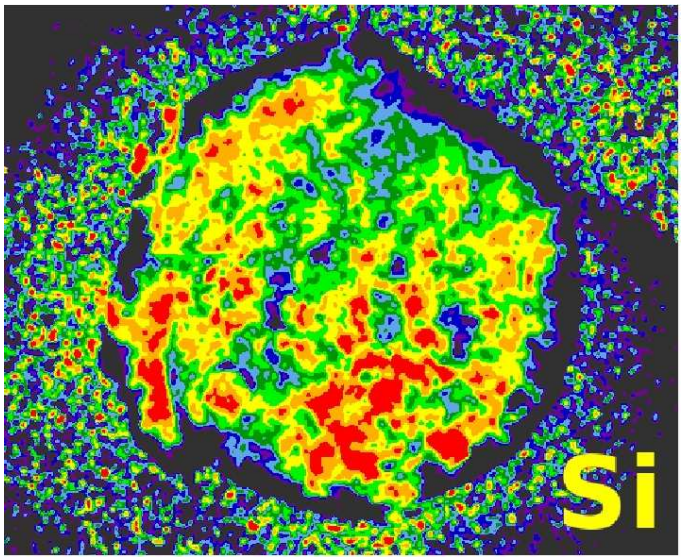}
\caption{Equivalent-width map for Si. The excess of line emission in
  the southern hemisphere, an indication of an asymmetry in the ejecta
  distribution, is approximately 5\% stronger than in the northern
  hemisphere.
\label{eqwidth}
}
\end{figure}

We can also examine the spatial distribution of the Si, though only in
the plane of the sky. In Figure~\ref{eqwidth}, we show an equivalent
width map of the Si-rich ejecta. We follow the procedure laid out in
\citet{winkler14}, where maps like this were created for SN 1006. The
equivalent width maps show the strength of the Si line at $\sim$1.8
keV relative to the strength of the underlying continuum emission and
serves as a proxy for the amount of Si present in a given spatial
region. We find that the spatial distribution is fairly uniform, with
only a slight increase in the line strength (at the $\sim$5\% level)
in the southern portion of the remnant. This is consistent with a
similar map shown in \citet{hwang02}, which used the ACIS-S array and
cut off the southernmost portion of the remnant. This is in contrast
to what \citet{winkler14} found in SN 1006, where the Si was
significantly stronger in the SE than in the NW.

\begin{figure}[h!]
\includegraphics[width=11cm]{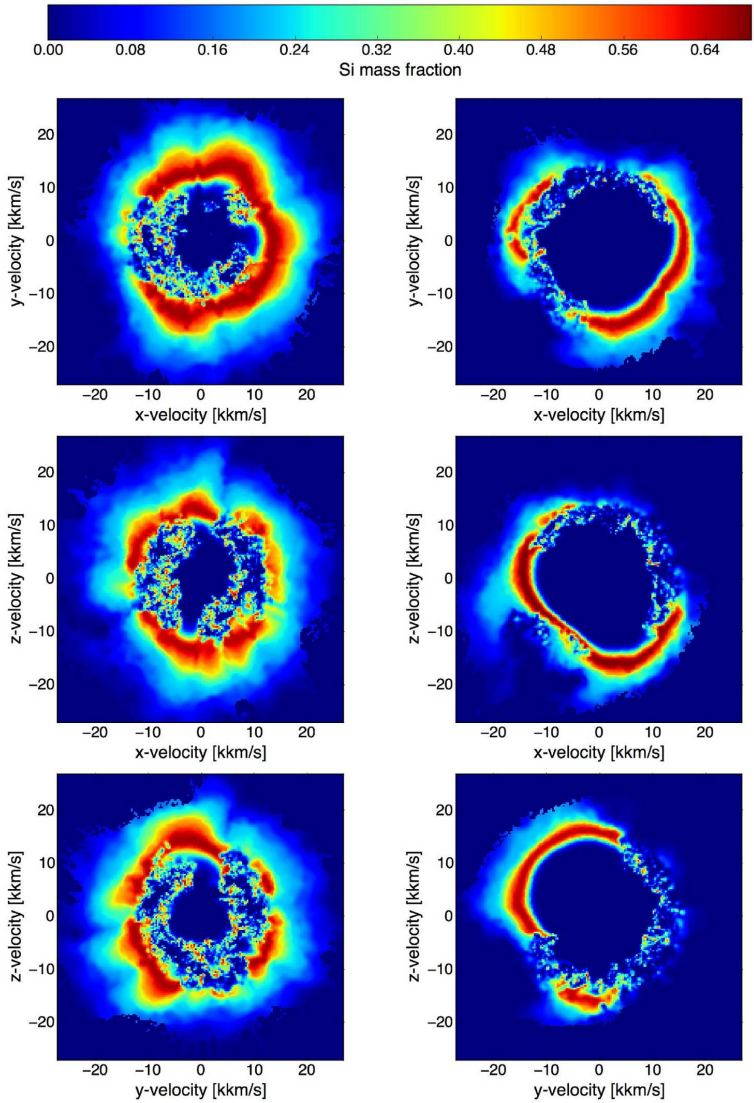}
\caption{Slices through all three coordinate axis planes for the
  velocity distribution of Si shortly after the explosion. The N100
  model of \citet{seitenzahl13} is shown on the left; the N5 model is
  on the right.
\label{velocity_dispersion}
}
\end{figure}

\citet{seitenzahl13} explored various explosion models for Type Ia
SNe. They compare three-dimensional hydrodynamical simulations for
various delayed-detonation models for the first $\sim 100$ seconds
post-explosion. They calculate models with varying ignition conditions
of the deflagration in the white dwarf at several values of the
central density. Among the many outputs of these models are the
velocity distributions of several elements, including Si. We do not,
yet, have end-to-end simulations from the explosion of a supernova all
the way to the supernova remnant phase hundreds of years
later. Nonetheless, we can compare the velocity distribution that we
see now to those produced in the explosion, as they are clearly
connected during the ejecta-dominated phase of the remnant.

In Figure~\ref{velocity_dispersion}, we show a comparison of
velocities in all three coordinate planes for both the ``N100'' and
``N5'' models, where the numbers represent the number of ignition
points within the white dwarf. We would like to emphasise here that
the number of ignition sparks should not be taken literally. In the
numerical simulations, the number and position of ignition sparks
serves as a means to control the rate of fuel consumption
(``deflagration strength'') and symmetry of the deflagration. For more
discussion, see \citet{sim13}. As can be seen from this figure and
from those in \citet{seitenzahl13}, models with more ignition kernels
lead to more symmetric explosions, both in the spatial distribution of
material and the velocity distribution. While we cannot conclusively
select only one of the models that best fit the observational data, we
conclude that the models with weakly ignited and asymmetric
deflagrations, such as the ``N3'' and ``N5'' models, are disfavored
for Tycho's SNR.


\begin{figure}[h!]
\includegraphics[width=15cm]{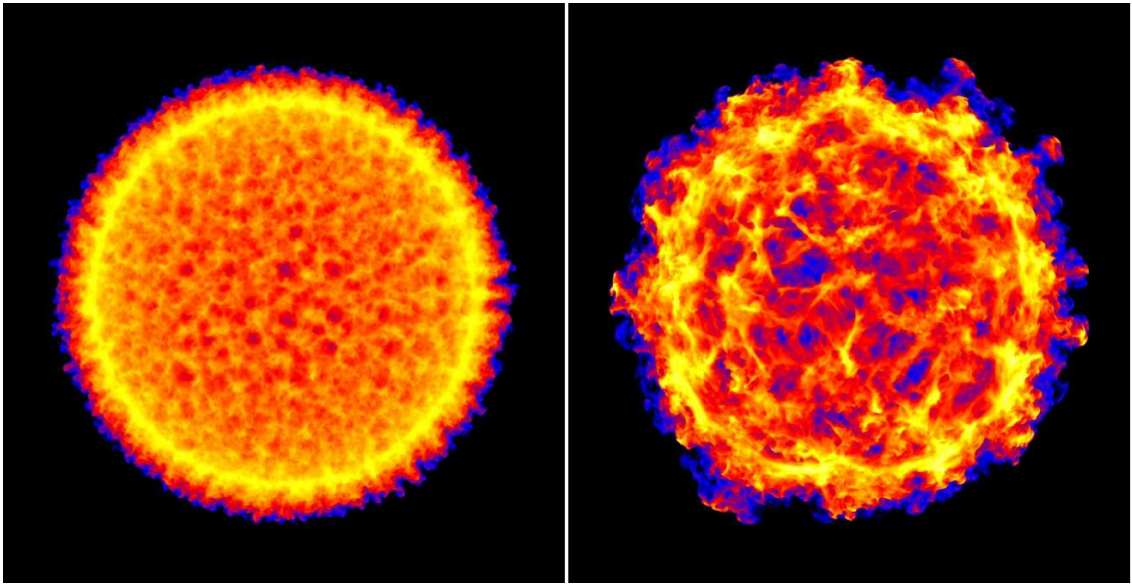}
\caption{Hydrodynamical simulations of Tycho at the current
  epoch. These are two-dimensional projections of a three-dimensional
  simulation, described in the text. The model on the left has a
  smooth initial ejecta density profile; the one on the right consists of
  clumpy initial ejecta. The images are on the same spatial scale.
\label{smoothvsclumpy}
}
\end{figure}

One of the more notable features of Tycho is the presence of ejecta
knots that protrude in front of the forward shock. These knots have
been discussed by many authors before. \citet{wang01} cite these knots
as evidence of initial clumping in the ejecta, saying that they cannot be
reproduced by fluid instabilities in a smooth ejecta profile in
two-dimensional hydrodynamical simulations. \citet{orlando12} reached
a similar conclusion using three-dimensional simulations regarding the
necessity of clumping in the ejecta. By contrast, \citet{warren13}
found that by varying the adiabatic index of the shocked gas,
$\gamma$, their three-dimensional hydro simulations could reproduce
the presence of ejecta knots ahead of the forward shock using smooth
ejecta without any initial clumpiness. They concluded that clumpiness
is not a necessary condition to explain the morphology of Tycho.

These and other previous studies focused on modeling the morphology of
the remnant. We now have the ability to go one step further and
compare the dynamics of Tycho with those from simulated data. We
compare our observed spatial velocities with two hydrodynamical
models: a smooth initial ejecta profile (which develops a "clumpy" structure 
over time due to fluid instabilities) and a clumpy initial profile. These simulations
are described in detail in \citet{warren13}; briefly, they use the
hydrodynamics code VH-1 on a yin-yang spherical overset grid with 0.25
degree angular resolution. The model assumes the exponential density
profile of \citet{dwarkadas98} and ``standard'' explosion parameters
of 10$^{51}$ ergs with 1.4 $\msun$ of ejecta. The clumpy model is
produced using a Perlin algorithm \citep{perlin85} to generate noise with a maximum
angular scale of $\sim 20$ degrees and a maximum to minimum density
contrast of 6.

Roughly a half-dozen knots protrude beyond Tycho's forward shock. To
remove the uncertainty in the absolute velocity caused by the unknown
distance to Tycho, we use the dimensionless deceleration parameter,
$m$ ($\equiv vt/R$), as our method of comparison. Here, $v$ is the proper motion (in arcseconds per year), $t$ is the time since explosion, and $R$ is the distance from the explosion site in arcseconds, which we assume to be the site determined in Paper II (slightly offset from the geometric center of the remnant). We compare motion only in the plane of the sky, as we cannot be certain of the radius of
the knots in the Z-direction. We show the simulated images for both
the smooth and clumpy models, showing only the reverse-shocked ejecta,
in Figure~\ref{smoothvsclumpy}.

As can be seen from Table~\ref{velocities}, the $m$ values vary throughout the 
remnant as measured from the {\it Chandra} images, from as low as $\sim 0.3$ to as high as $\sim 0.9$, though the majority fall between 0.45-0.7. We measured the deceleration parameters for ejecta
knots in both the smooth and clumpy models, finding a spread in values for the smooth model from $\sim 0.4$ to 0.6 and for the clumpy model from $\sim 0.5$ to 0.7. The dispersion in our measured
values for Tycho is large enough to accommodate both models. We
caution that care should be taken in comparing the observations to the
models here, as we only have two hydrodynamic models. Large variations
may exist between various clumpy ejecta models for supernova remnants.


Future work should further explore various ejecta distributions via
multi-dimensional hydrodynamical simulations. For example, our work
here only applies to the Si-rich ejecta, but \citet{wang01} suggest
that the level of clumpiness may differ between the Si and Fe-rich
ejecta. Observationally, studies such as this are possible for other
remnants, though the observational times required for {\it Chandra}
are substantial. Future X-ray missions with substantially increased
effective area will significantly reduce the observing time
requirements, but we stress that for this type of science, this must
be coupled with high-resolution imaging. Such missions would also
allow this analysis to be done for other ejecta species. Ar, Ca, and
Fe K$\alpha$ are all present in Tycho, but are too faint to do using
the existing {\it Chandra} observations. Finally, young remnants like
Tycho should continue to be observed regularly with {\it Chandra},
since measurement uncertainties will decrease with time as the
proper motions get larger.

\section{Conclusions}

Young supernova remnants offer a somewhat rare opportunity in
astronomy: the chance to observe spatial evolution in real time. {\it
  Chandra} X-ray observations of Tycho's supernova remnant spread out
over 12 years constitute several percent of the lifetime of the
remnant, enough to measure the expansion of the ejecta knots via their
proper motion. Spectroscopic analysis yields the line-of-sight
velocity, giving the first three-dimensional velocity map for the
remnant of a Type Ia supernova.

We have nearly five dozen knots of ejecta for which we can reliably
measure the proper motion in the plane of the sky and the
red/blueshift of the spectral lines of Si and S. We find no measurable asymmetry 
in the velocity of the Si-rich ejecta in any direction. Some models of Type Ia SNe 
predict that such a velocity asymmetry should exist. When we compare our observations
with the delayed detonation models of \citet{seitenzahl13}, we favor
models with strongly ignited, symmetric deflagrations, such as the
N100 model.

We see very little spatial asymmetry when looking at the Si-rich
ejecta. Equivalent-width maps of the $\sim 1.8$ keV line of Si show
that the Si is quite homogeneous. This is in contrast to some other
young Type Ia remnants, such as SN 1006 and G1.9+0.3. The deceleration parameters
we measure for the ejecta knots in the plane of the sky are consistent with 
hydrodynamical simulations of both smooth and clumpy ejecta profiles. 

\acknowledgements

We thank Paul Plucinsky and R. Nick Durham at the {\it Chandra X-ray
  Center} for their assistance in quantifying the ACIS gain
calibration for our spectral analysis. Support for this work was
provided through Chandra Award GO4-15074Z issued by the Chandra X-ray
Observatory Center, which is operated by the Smithsonian Astrophysical
Observatory for and on behalf of NASA under contract NAS8-03060.

\begin{deluxetable}{lccccccccc}
\tablecolumns{10} 
\tablewidth{0pc} 
\tabletypesize{\footnotesize}
\tablecaption{Velocity Measurements for Ejecta Knots} \tablehead{ \colhead{Region} & R.A. & Dec. & $\mu_{X}$ & $\mu_{Y}$ & V$_{X}$ & V$_{Y}$ & V$_{Z}$ & V$_{Total}$ & $m$}

\startdata

1 & 6.179 & 64.146 & 0.307$_{0.294}^{0.319}$ & 0.014$_{0.013}^{0.015}$ & 5090$_{4880}^{5300}$ & 230$_{220}^{240}$ & 620$_{450}^{780}$ & 5140 & 0.56\\
2 & 6.181 & 64.161 & 0.301$_{0.292}^{0.310}$ & 0.060$_{0.058}^{0.061}$ & 4990$_{4840}^{5140}$ & 990$_{960}^{1020}$ & 710$_{420}^{990}$ & 5140 & 0.54\\
3 & 6.187 & 64.158 & 0.320$_{0.283}^{0.356}$ & 0.097$_{0.086}^{0.108}$ & 5310$_{4700}^{5910}$ & 1610$_{1430}^{1800}$ & 1880$_{1510}^{2230}$ & 5860 & 0.73\\
4 & 6.196 & 64.153 & 0.236$_{0.223}^{0.248}$ & 0.066$_{0.063}^{0.070}$ & 3910$_{3700}^{4120}$ & 1100$_{1040}^{1160}$ & 880$_{640}^{1130}$ & 4160 & 0.41\\
5 & 6.198 & 64.128 & 0.256$_{0.233}^{0.278}$ & -0.108$_{-0.098}^{-0.117}$ & 4240$_{3860}^{4610}$ & -1790$_{-1630}^{-1950}$ & 1140$_{920}^{1360}$ & 4740 & 0.38\\
6 & 6.201 & 64.107 & 0.240$_{0.222}^{0.257}$ & -0.152$_{-0.140}^{-0.163}$ & 3990$_{3680}^{4270}$ & -2520$_{-2330}^{-2700}$ & -660$_{-520}^{-810}$ & 4760 & 0.45\\
7 & 6.206 & 64.131 & 0.183$_{0.169}^{0.197}$ & -0.144$_{-0.132}^{-0.155}$ & 3040$_{2800}^{3280}$ & -2380$_{-2190}^{-2570}$ & -640$_{-430}^{-850}$ & 3920 & 0.56\\
8 & 6.212 & 64.154 & 0.205$_{0.190}^{0.219}$ & 0.035$_{0.033}^{0.038}$ & 3400$_{3160}^{3640}$ & 590$_{550}^{630}$ & 1790$_{1560}^{1990}$ & 3890 & 0.45\\
9 & 6.215 & 64.110 & 0.260$_{0.250}^{0.270}$ & -0.094$_{-0.091}^{-0.098}$ & 4320$_{4160}^{4490}$ & -1560$_{1500}^{-1620}$ & -1870$_{-1480}^{-2170}$ & 4960 & 0.55\\
10 & 6.218 & 64.139 & 0.210$_{0.191}^{0.229}$ & -0.038$_{-0.035}^{-0.042}$ & 3480$_{3170}^{3800}$ & -630$_{-580}^{-690}$ & -420$_{-330}^{-500}$ & 3560 & 0.47\\
11 & 6.222 & 64.106 & 0.202$_{0.191}^{0.214}$ & -0.161$_{-0.152}^{-0.170}$ & 3360$_{3160}^{3550}$ & -2690$_{-2510}^{-2820}$ & -660$_{-530}^{-780}$ & 4340 & 0.44\\
12 & 6.247 & 64.177 & 0.120$_{0.101}^{0.141}$ & 0.132$_{0.111}^{0.155}$ & 1990$_{1680}^{2350}$ & 2190$_{1850}^{2580}$ & -1180$_{-900}^{-1430}$ & 3190 & 0.29\\
13 & 6.248 & 64.085 & 0.175$_{0.146}^{0.210}$ & -0.224$_{-0.187}^{-0.270}$ & 2900$_{2420}^{3490}$ & -3720$_{-3110}^{-4480}$ & 1490$_{1310}^{1830}$ & 4950 & 0.62\\
14 & 6.249 & 64.165 & 0.085$_{0.065}^{0.104}$ & 0.108$_{0.083}^{0.131}$ & 1410$_{1080}^{1720}$ & 1790$_{1370}^{2180}$ & 2600$_{2340}^{2890}$ & 3460 & 0.33\\
15 & 6.253 & 64.109 & 0.185$_{0.175}^{0.194}$ & -0.138$_{-0.131}^{-0.145}$ & 3060$_{2900}^{3220}$ & -2290$_{-2170}^{-2410}$ & 2120$_{1840}^{2400}$ & 4370 & 0.62\\
16 & 6.260 & 64.139 & 0.135$_{0.108}^{0.160}$ & 0.003$_{0.002}^{0.004}$ & 2240$_{1800}^{2660}$ & 50$_{40}^{60}$ & 770$_{420}^{1130}$ & 2370 & 0.43\\
17 & 6.262 & 64.137 & 0.118$_{0.090}^{0.144}$ & -0.020$_{-0.016}^{-0.025}$ & 1950$_{1500}^{2390}$ & -340$_{-260}^{-410}$ & 2950$_{2700}^{3180}$ & 3560 & 0.37\\
18 & 6.268 & 64.131 & 0.114$_{0.128}^{0.155}$ & 0.179$_{0.149}^{0.207}$ & 1890$_{1580}^{2190}$ & 2970$_{2480}^{3430}$ & -1930$_{-1680}^{-2170}$ & 4020 & 0.44\\
19 & 6.274 & 64.130 & 0.141$_{0.128}^{0.155}$ & -0.050$_{-0.045}^{-0.054}$ & 2350$_{2130}^{2570}$ & -820$_{-740}^{-900}$ & -1110$_{-930}^{-1290}$ & 2730 & 0.58\\
20 & 6.281 & 64.070 & 0.070$_{0.068}^{0.071}$ & -0.358$_{-0.347}^{-0.367}$ & 1150$_{1120}^{1190}$ & -5930$_{-5760}^{-6090}$ & -1060$_{-850}^{-1280}$ & 6140 & 0.58\\
21 & 6.282 & 64.170 & 0.055$_{0.049}^{0.060}$ & 0.117$_{0.106}^{0.128}$ & 910$_{820}^{990}$ & 1950$_{1760}^{2120}$ & -2150$_{-1730}^{-2940}$ & 3040 & 0.41\\
22 & 6.294 & 64.067 & 0.028$_{0.027}^{0.028}$ & -0.369$_{-0.360}^{-0.379}$ & 460$_{450}^{470}$ & -6130$_{-5970}^{-6280}$ & -920$_{-770}^{-1070}$ & 6220 & 0.58\\
23 & 6.294 & 64.095 & 0.003$_{0.002}^{0.005}$ & -0.095$_{-0.064}^{-0.128}$ & 60$_{40}^{80}$ & -1580$_{-1060}^{-2130}$ & 2130$_{1940}^{2330}$ & 2660 & 0.81\\
24* & 6.294 & 64.197 & 0.037$_{0.034}^{0.040}$ & 0.169$_{0.157}^{0.181}$ & 620$_{570}^{660}$ & 2800$_{2600}^{3000}$ & 3190$_{2790}^{3390}$ & 4290 & 0.46\\
25 & 6.299 & 64.082 & 0.073$_{0.069}^{0.077}$ & -0.243$_{-0.230}^{-0.256}$ & 1210$_{1150}^{1270}$ & -4040$_{-3820}^{-4240}$ & -320$_{-130}^{-490}$ & 4230 & 0.64\\
26* & 6.299 & 64.192 & 0.011$_{0.008}^{0.013}$ & 0.158$_{0.117}^{0.186}$ & 180$_{130}^{210}$ & 2610$_{1950}^{3090}$ & -880$_{-610}^{-1140}$ & 2760 & 0.27\\
27* & 6.307 & 64.198 & -0.005$_{-0.005}^{-0.006}$ & 0.246$_{0.228}^{0.263}$ & -90$_{-80}^{-90}$ & 4080$_{3780}^{4370}$ & 1950$_{1650}^{2250}$ & 4520 & 0.62\\
28 & 6.318 & 64.169 & 0.018$_{0.016}^{0.019}$ & 0.179$_{0.166}^{0.192}$ & 290$_{270}^{310}$ & 2970$_{2760}^{3180}$ & 3360$_{2570}^{3660}$ & 4490 & 0.71\\
29 & 6.322 & 64.180 & -0.007$_{-0.006}^{-0.009}$ & 0.153$_{0.128}^{0.177}$ & -120$_{-100}^{-140}$ & 2540$_{2130}^{2940}$ & -1380$_{-1130}^{-1640}$ & 2890 & 0.45\\
30* & 6.327 & 64.213 & 0.005$_{0.005}^{0.005}$ & 0.393$_{0.383}^{0.402}$ & 80$_{80}^{80}$ & 6520$_{6360}^{6670}$ & 770$_{90}^{1540}$ & 6560 & 0.62\\
31 & 6.329 & 64.187 & -0.027$_{-0.024}^{-0.031}$ & 0.161$_{0.139}^{0.182}$ & -450$_{-390}^{-510}$ & 2680$_{2310}^{3020}$ & -2630$_{-2260}^{-2990}$ & 3780 & 0.44\\
32 & 6.342 & 64.074 & -0.026$_{-0.025}^{-0.027}$ & -0.292$_{-0.281}^{-0.303}$ & -430$_{-420}^{-450}$ & -4840$_{4660}^{-5030}$ & 410$_{200}^{620}$ & 4880 & 0.52\\
33 & 6.344 & 64.087 & -0.032$_{-0.029}^{-0.036}$ & -0.233$_{-0.210}^{-0.257}$ & -540$_{-480}^{-590}$ & -3870$_{-3480}^{-4270}$ & -1120$_{-950}^{-1300}$ & 4070 & 0.57\\
34 & 6.344 & 64.095 & -0.053$_{-0.046}^{-0.060}$ & -0.202$_{-0.177}^{-0.229}$ & -870$_{-770}^{-990}$ & -3350$_{-2940}^{-3800}$ & 1920$_{1630}^{2190}$ & 3960 & 0.62\\
35 & 6.360 & 64.198 & -0.058$_{-0.054}^{-0.062}$ & 0.310$_{0.291}^{0.319}$ & -960$_{-900}^{-1030}$ & 5140$_{4820}^{5480}$ & 1140$_{910}^{1380}$ & 5360 & 0.82\\
36 & 6.370 & 64.114 & -0.084$_{-0.076}^{-0.093}$ & -0.108$_{-0.097}^{-0.118}$ & -1400$_{-1260}^{-1540}$ & -1790$_{-1610}^{-1970}$ & -1310$_{-1060}^{-1540}$ & 2620 & 0.68\\
37 & 6.371 & 64.195 & -0.097$_{-0.093}^{-0.102}$ & 0.305$_{0.291}^{0.319}$ & -1620$_{-1540}^{-1690}$ & 5060$_{4830}^{5290}$ & 2030$_{1810}^{2260}$ & 5690 & 0.71\\
38 & 6.375 & 64.177 & -0.119$_{-0.111}^{-0.128}$ & 0.146$_{0.136}^{0.157}$ & -1980$_{-1850}^{-2120}$ & 2430$_{2260}^{2610}$ & 2040$_{1770}^{2310}$ & 3740 & 0.50\\
39 & 6.396 & 64.088 & -0.145$_{-0.138}^{-0.152}$ & -0.246$_{-0.234}^{-0.258}$ & -2410$_{-2290}^{-2530}$ & -4080$_{-3880}^{-4290}$ & 1000$_{690}^{1370}$ & 4840 & 0.67\\
40 & 6.399 & 64.173 & -0.182$_{-0.168}^{-0.197}$ & 0.100$_{0.092}^{0.108}$ & -3030$_{-2780}^{-3270}$ & 1660$_{1530}^{1800}$ & -1410$_{-1210}^{-1600}$ & 3730 & 0.63\\
41 & 6.400 & 64.165 & -0.201$_{-0.183}^{-0.218}$ & 0.080$_{0.073}^{0.087}$ & -3330$_{-3040}^{-3610}$ & 1330$_{1210}^{1440}$ & -2630$_{-2370}^{-2890}$ & 4450 & 0.94\\
42 & 6.403 & 64.193 & -0.088$_{-0.082}^{-0.094}$ & 0.263$_{0.244}^{0.281}$ & -1460$_{-1360}^{-1560}$ & 4370$_{4060}^{4660}$ & -310$_{-160}^{-460}$ & 4620 & 0.48\\
43 & 6.422 & 64.186 & -0.158$_{-0.152}^{-0.165}$ & 0.128$_{0.123}^{0.133}$ & -2620$_{-2520}^{-2740}$ & 2120$_{2030}^{2210}$ & 1000$_{880}^{1130}$ & 3520 & 0.58\\
44 & 6.431 & 64.107 & -0.292$_{-0.273}^{-0.309}$ & -0.058$_{-0.054}^{-0.061}$ & -4840$_{-4530}^{-5120}$ & -960$_{-900}^{-1010}$ & -1380$_{-1050}^{-1850}$ & 5120 & 0.65\\
45 & 6.433 & 64.176 & -0.242$_{-0.220}^{-0.262}$ & 0.133$_{0.121}^{0.144}$ & -4010$_{-3660}^{-4350}$ & 2200$_{2010}^{2390}$ & 680$_{430}^{940}$ & 4620 & 0.43\\
46 & 6.434 & 64.161 & -0.165$_{-0.156}^{-0.175}$ & 0.126$_{0.119}^{0.134}$ & -2740$_{-2590}^{-2900}$ & 2100$_{1980}^{2220}$ & -1260$_{-1120}^{-1390}$ & 3680 & 0.84\\
47 & 6.447 & 64.188 & -0.186$_{-0.182}^{-0.191}$ & 0.164$_{0.160}^{0.168}$ & -3080$_{-3020}^{-3160}$ & 2720$_{2660}^{2790}$ & 480$_{320}^{630}$ & 4140 & 0.55\\
48 & 6.452 & 64.178 & -0.273$_{-0.264}^{-0.282}$ & 0.162$_{0.157}^{0.168}$ & -4540$_{-4380}^{-4680}$ & 2690$_{2600}^{2780}$ & 640$_{500}^{790}$ & 5310 & 0.62\\
49 & 6.463 & 64.146 & -0.214$_{-0.193}^{-0.235}$ & -0.018$_{-0.016}^{-0.019}$ & -3560$_{-3200}^{-3890}$ & -290$_{-260}^{-320}$ & -1040$_{-740}^{-1330}$ & 3720 & 0.61\\
50 & 6.465 & 64.095 & -0.254$_{-0.245}^{-0.264}$ & -0.223$_{-0.214}^{-0.231}$ & -4220$_{-4070}^{-4380}$ & -3700$_{-3560}^{-3830}$ & 1730$_{1500}^{1910}$ & 5870 & 0.75\\
51 & 6.471 & 64.162 & -0.228$_{-0.223}^{-0.233}$ & 0.152$_{0.148}^{0.155}$ & -3780$_{-3710}^{-3860}$ & 2510$_{2460}^{2570}$ & 390$_{100}^{690}$ & 4560 & 0.61\\
52 & 6.471 & 64.173 & -0.246$_{-0.240}^{-0.252}$ & 0.127$_{0.123}^{0.130}$ & -4090$_{-3980}^{-4190}$ & 2100$_{2040}^{2150}$ & 610$_{410}^{810}$ & 4630 & 0.56\\
53 & 6.476 & 64.136 & -0.317$_{-0.311}^{-0.323}$ & 0.023$_{0.023}^{0.024}$ & -5260$_{-5170}^{-5350}$ & 390$_{380}^{390}$ & -980$_{-550}^{-1430}$ & 5360 & 0.65\\
54 & 6.477 & 64.104 & -0.304$_{-0.296}^{-0.310}$ & -0.185$_{-0.181}^{-0.189}$ & -5040$_{-4910}^{-5150}$ & -3070$_{-3000}^{-3140}$ & 880$_{230}^{1500}$ & 5970 & 0.72\\
55 & 6.489 & 64.126 & -0.330$_{-0.327}^{-0.335}$ & -0.040$_{-0.040}^{-0.041}$ & -5480$_{-5420}^{-5560}$ & -660$_{-660}^{-670}$ & -2460$_{-2290}^{-2610}$ & 6040 & 0.64\\
56 & 6.490 & 64.131 & -0.279$_{-0.273}^{-0.284}$ & 0.076$_{0.074}^{0.077}$ & -4620$_{-4530}^{-4710}$ & 1260$_{1230}^{1280}$ & -1390$_{-1190}^{-1600}$ & 4990 & 0.48\\
57 & 6.492 & 64.112 & -0.360$_{-0.354}^{-0.367}$ & -0.079$_{-0.077}^{-0.080}$ & -5980$_{-5880}^{-6090}$ & -1300$_{-1280}^{-1330}$ & -1600$_{-1200}^{-2130}$ & 6320 & 0.62\\

\enddata

\tablecomments{Region number corresponds to numbers shown on
  Figure~\ref{regions}. Regions are numbered in order of ascending
  right ascension, and were drawn on the 2009 observation. R.A. and
  Dec. are given in decimal degrees in J2000 coordinates. All
  velocities given in km s$^{-1}$ and rounded to the nearest 10 km s$^{-1}$. X is positive to the west, Y is
  positive to the north, and Z is positive away from the
  observer. Uncertainties are statistical only. Systematic
  uncertainties due to WCS alignment for V$_{X}$ and V$_{Y}$ are
  negligible. Systematic uncertainties for V$_{Z}$ due to CCD gain
  calibration are 900 km s$^{-1}$ for all regions except those with an
  asterisk, where the systematic uncertainties are 3000 km
  s$^{-1}$. Deceleration parameter, $m$, is described in the text and
  only measured in the plane of the sky, assuming the calculated explosion
  center determined in Paper II of $\alpha$ =
0$^{h}$25$^{m}$22.6$^{s}$ and $\delta$ = 64$^{\circ}$8$'$32.7$''$.}
\label{velocities}
\end{deluxetable}

\begin{deluxetable}{lccc}
\tablecolumns{4} 
\tablewidth{0pc} 
\tabletypesize{\footnotesize}
\tablecaption{XSPEC NEI Model Fits to Ejecta Knot Spectra} \tablehead{ \colhead{Region} & $\tau_{i}$ (10$^{10}$ cm$^{-3}$ s) & kT (keV) & $\chi^{2}$/d.o.f.}

\startdata

1 & 5.36 & 1.23 & 135/99\\
2 & 2.12 & 1.89 & 139/99\\
3 & 6.29 & 1.26 & 97/98\\
4 & 6.54 & 1.13 & 128/94\\
5 & 1.96 & 1.54 & 153/100\\
6 & 4.29 & 1.56 & 124/99\\
7 & 4.41 & 1.65 & 189/106\\
8 & 6.51 & 1.39 & 188/99\\
9 & 5.13 & 1.43 & 110/93\\
10 & 8.56 & 1.17 & 217/103\\
11 & 7.57 & 1.10 & 225/105\\
12 & 2.63 & 2.61 & 165/100\\
13 & 5.09 & 1.35 & 149/94\\
14 & 11.0 & 1.35 & 149/94\\
15 & 13.2 & 0.94 & 114/92\\
16 & 5.76 & 1.37 & 107/92\\
17 & 8.01 & 1.20 & 185/96\\
18 & 6.91 & 1.11 & 158/98\\
19 & 19.2 & 0.83 & 252/93\\
20 & 5.11 & 1.15 & 104/93\\
21 & 3.98 & 1.13 & 100/93\\
22 & 4.31 & 1.29 & 103/98\\
23 & 8.29 & 1.12 & 204/98\\
24 & 7.83 & 1.50 & 151/102\\
25 & 12.7 & 0.88 & 145/100\\
26 & 3.92 & 2.36 & 159/101\\
27 & 9.97 & 1.21 & 238/101\\
28 & 5.27 & 1.21 & 135/93\\
29 & 6.68 & 1.26 & 95/96\\
30 & 5.05 & 2.28 & 70/73\\
31 & 5.80 & 1.39 & 108/96\\
32 & 4.93 & 1.20 & 97/98\\
33 & 11.4 & 0.90 & 123/93\\
34 & 18.7 & 0.88 & 205/93\\
35 & 6.22 & 1.24 & 159/98\\
36 & 6.87 & 1.12 & 108/95\\
37 & 8.18 & 1.19 & 147/99\\
38 & 25.8 & 0.84 & 181/104\\
39 & 2.23 & 2.10 & 109/100\\
40 & 10.1 & 1.03 & 163/96\\
41 & 8.98 & 0.95 & 96/95\\
42 & 5.48 & 1.42 & 144/100\\
43 & 9.79 & 1.06 & 244/100\\
44 & 8.64 & 0.98 & 99/84\\
45 & 4.90 & 1.51 & 113/94\\
46 & 7.76 & 1.11 & 119/98\\
47 & 8.12 & 1.05 & 157/98\\
48 & 9.04 & 0.98 & 148/98\\
49 & 13.5 & 0.93 & 129/90\\
50 & 5.59 & 1.20 & 141/100\\
51 & 4.72 & 1.36 & 76/92\\
52 & 6.95 & 1.02 & 163/94\\
53 & 8.35 & 1.04 & 123/100\\
54 & 104 & 0.57 & 102/100\\
55 & 4.45 & 1.47 & 310/100\\
56 & 4.94 & 1.38 & 119/90\\
57 & 8.43 & 0.86 & 204/80\\

\enddata

\tablecomments{$\tau_{i}$ is ionization timescale, the integral of
  electron density over time in the post-shock gas.}
\label{neifits}
\end{deluxetable}


\begin{thebibliography}{}

\bibitem[Albinson et al.(1986)]{albinson86}
Albinson, J.S., Tuffs, R.J., Swinbank, E., \& Gull, S.F. 1986, MNRAS, 219, 427

\bibitem[Baade(1945)]{baade45}
Baade, W. 1945, ApJ, 102, 309

\bibitem[Badenes et al.(2006)]{badenes06}
Badenes, C., Borkowski, K.J., Hughes, J.P., Hwang, U., \& Bravo, E. 2006, ApJ, 645, 1373

\bibitem[Borkowski et al.(2013)]{borkowski13}
Borkowski, K.J. et al. 2013, ApJ, 771, 9

\bibitem[Borkowski et al.(2017)]{borkowski17}
Borkowski, K.J., Gwynne, P., Reynolds, S.P., Green, D.A., Hwang, U., Petre, R., \& Willett, R., 2017, accepted

\bibitem[Cassam-Chena\"{i} et al.(2007)]{cassamchenai07}
Cassam-Chena\"{i}, G., Hughes, J.P., Ballet, J., \& Decourchelle, A. 2007, ApJ, 665, 315

\bibitem[Chevalier et al.(1980)]{chevalier80}
Chevalier, R.A., Kirshner, R.P., \& Raymond, J.C. 1980, ApJ, 235, 186

\bibitem[Delaney et al.(2010)]{delaney10}
Delaney, T., et al. 2010, ApJ, 725, 2038

\bibitem[Dwarkadas \& Chevalier(1998)]{dwarkadas98}
Dwarkadas, V.V. \& Chevalier, R.A. 1998, ApJ, 497, 807

\bibitem[Fesen et al.(2006)]{fesen06}
Fesen, R.A., et al. 2006, ApJ, 645, 283

\bibitem[Furuzawa et al.(2009)]{furuzawa09}
Furuzawa, A., et al. 2009, ApJ, 693, 61

\bibitem[Hayato et al.(2010)]{hayato10}
Hayato, A., et al. 2010, ApJ, 725, 894

\bibitem[Hughes(2000)]{hughes00}
Hughes, J.P. 2000, ApJ, 545, 53

\bibitem[Hwang et al.(2002)]{hwang02}
Hwang, U., Decourchelle, A., Holt, S., Petre, R. 2002, ApJ, 581, 1101

\bibitem[Katsuda et al.(2008)]{katsuda08}
Katsuda, S., Tsunemi, H., \& Mori, K. 2008, ApJ, 678, 35

\bibitem[Katsuda et al.(2010)]{katsuda10} Katsuda, S., Petre, R.,
  Hughes, J.P., Hwang, U., Yamaguchi, H., Hayato, A., Mori, K.,
  Tsunemi, H. 2010, ApJ, 709, 1387

\bibitem[Kerzendorf et al.(2013)]{kerzendorf13}
Kerzendorf, W.E., et al. 2013, ApJ, 774, 99

\bibitem[Krause et al.(2008)]{krause08}
Krause, O., Tanaka, M., Usuda, T., Hattori, T., Goto, M., Birkmann, S., \& Nomoto, K. 2008, {\it Nature}, 456, 617

\bibitem[Maeda et al.(2010)]{maeda10}
Maeda, K., et al. 2010, Nature, 466, 82

\bibitem[Motohara et al.(2006)]{motohara06}
Motohara, K., et al. 2006, ApJ, 652, 101

\bibitem[Orlando et al.(2012)]{orlando12}
Orlando, S., Bocchino, F., Miceli, M., Petruk, O., \& Pumo, M.L. 2012, ApJ, 749, 156

\bibitem[Perlin(1985)]{perlin85}
Perlin, K. 1985, SIGGRAPH Comput. Graph. 19, 287

\bibitem[Ressler et al.(2014)]{ressler14}
Ressler, S.M., Katsuda, S., Reynolds, S.P., Long, K.S., Petre, R., Williams, B.J., \& Winkler, P.F. 2014, ApJ, 790, 85

\bibitem[Rest et al.(2008)]{rest08}
Rest, A., et al. 2008, ApJ, 681, 81

\bibitem[Reynolds \& Keohane(1999)]{reynolds99}
Reynolds, S.P. \& Keohane, J.W. 1999, ApJ, 525, 368

\bibitem[Reynoso et al.(1997)]{reynoso97}
Reynoso, E.M., Moffett, D.A., Goss, W.M., Dubner, G.M., Dickel, J.R., Reynolds, S.P., \& Giacani, E.B. 1997, ApJ, 491, 816

\bibitem[Ruiz-Lapuente et al.(2004)]{ruiz04}
Ruiz-Lapuente, P., et al. 2004, Nature, 431, 1069

\bibitem[Sato \& Hughes(2016)]{sato16}
Sato, T. \& Hughes, J.P., arxiv:1605.09059

\bibitem[Schaefer \& Pagnotta(2012)]{schaefer12}
Schaefer, B.E. \& Pagnotta, A. 2012, Nature, 481, 164

\bibitem[Schwarz et al.(1995)]{schwarz95}
Schwarz, U.J., Goss, W.M., Kalberla, P.M., \& Benaglia, P. 1995, A\&A, 299, 193

\bibitem[Seitenzahl et al.(2013)]{seitenzahl13}
Seitenzahl, I.R., et al. 2013, MNRAS, 429, 1156

\bibitem[Sim et al.(2013)]{sim13}
Sim, S.A., et al. 2013, MNRAS, 436, 333

\bibitem[Stephenson \& Green(2002)]{stephenson02}
Stephenson, F.R. \& Green, D.A. 2002, {\it Historical Supernovae and their Remnants}, Oxford University Press

\bibitem[Tran et al.(2015)]{tran15}
Tran, A., Williams, B.J., Petre, R., Ressler, S.M., \& Reynolds, S.P. 2015, ApJ, 812, 101

\bibitem[Wang \& Chevalier(2001)]{wang01}
Wang, C.-H. \& Chevalier, R.A. 2001, ApJ, 549, 1119

\bibitem[Wang \& Wheeler(2008)]{wang08}
Wang, L. \& Wheeler, J.C. 2008, ARA\&A, 46, 433

\bibitem[Warren et al.(2005)]{warren05}
Warren, J.S., et al. 2005, ApJ, 634, 376

\bibitem[Warren \& Blondin(2013)]{warren13}
Warren, D.C. \& Blondin, J.M. 2013, MNRAS, 429, 3099

\bibitem[Webbink(1984)]{webbink84}
Webbink, R.F. 1984, ApJ, 277, 355

\bibitem[Whelan \& Iben(1973)]{whelan73}
Whelan, J., \& Iben, I., Jr. 1973, ApJ, 186, 1007

\bibitem[Williams et al.(2013)]{williams13}
Williams, B.J., et al. 2013, ApJ, 770, 129

\bibitem[Williams et al.(2016)]{williams16}
Williams, B.J., et al. 2016, ApJL, 823, 32

\bibitem[Winkler et al.(2014)]{winkler14}
Winkler, P.F., et al. 2014, ApJ, 781, 65

\bibitem[Xue \& Schaefer(2015)]{xue15}
Xue, Z. \& Schaefer, B.E. 2015, ApJ, 809, 183

\bibitem[Yamaguchi et al.(2016)]{yamaguchi16}
Yamaguchi, H., Katsuda, S., \& Castro, D. et al. 2016, ApJ, 820, 3

\bibitem[Yamaguchi et al.(2017)]{yamaguchi17}
Yamaguchi, H., Hughes, J.P., Badenes, C., Bravo, E., Seitenzahl, I.R., Martinez-Rodriguez, H., Park, S., \& Petre, R. 2017, ApJ, 834, 124

\end{thebibliography}
\end{document}